\documentclass[fleqn,usenatbib]{mnras}

\usepackage{newtxtext,newtxmath}
\usepackage[utf8]{inputenc}
\usepackage[T1]{fontenc}
\defcitealias{smi04}{Smith 2004}
\defcitealias{smi18}{Smith et al. 2018}
\defcitealias{bor19}{Bordiu \& Rizzo 2019}
\defcitealias{rem13}{Remmer et al.  2013}
\defcitealias{abr14}{Paper~I}
\defcitealias{hum99}{Humphreys et al. 1999}
\defcitealias{dor04}{Dorland et al. 2004}
\defcitealias{dav97}{Davidson et al. 1997}
\defcitealias{art11}{Artugau et al. 2011}
\defcitealias{ver05}{Verner et al.  2005}
\defcitealias{meh10}{Mehner et al. 2010}
\usepackage{graphicx}	% Including figure files
\title[$\eta$ Car ALMA images]{$\eta$ Carinae: high angular resolution continuum, H30$\alpha$ and He30$\alpha$ ALMA images}

\author[Abraham et al.]{
Zulema Abraham,$^{1}$\thanks{E-mail: zulema.abraham@iag.usp.br}
Pedro. P. B. Beaklini,$^{1,2}$
Pierre Cox,$^{3}$
Diego Falceta-Gon\c calves$^{4}$
\newauthor
and 
Lars-\AA ke  Nyman$^{5}$
\\
$^{1}$Instituto de Astronomia, Geof\'isica e Ci\^encias Atmosf\'ericas, Universidade de S\~ao Paulo \\
Rua do Mat\~ao 1226, CEP 05508-090, S\~ao Paulo, Brazil\\
$^{2}$National Radio Astronomy Observatory, 1003 Lopezville Road, Socorro, NM 87801, USA\\
$^{3}$Institut d'Astrophysique de Paris, Sorbonne Universit\'e, UPMC Universit\'e Paris 6 and CNRS,\\ UMR 7095, 98bis boulevard Arago, 75014, Paris, France\\
$^{4}$Escola de Artes, Ci\^encias e Humanidades, Universidade de S\~ao Paulo,\\ Rua Arlindo Bettio 1000, CEP 03828-000, S\~ao Paulo, Brazil\\
$^{5}$European Southern Observatory, Alonso de C\'ordoba 3107, Vitacura, Chile}

\date{Accepted XXX. Received YYY; in original form ZZZ}

\pubyear{2020}

\begin{document}
\label{firstpage}
\pagerange{\pageref{firstpage}--\pageref{lastpage}}
\maketitle

\begin{abstract}
We present images of $\eta$ Carinae in the recombination lines  H30$\alpha$ and He30$\alpha$ and  the underlying continuum  with 50~mas resolution (110 AU), obtained with ALMA. 
For the first time, the 230 GHz continuum image is resolved into a compact core, coincident with the binary system position, and  a weaker extended structure to the NW of the compact source. 
Iso-velocity images of the H30$\alpha$ recombination line show at least 16 unresolved sources with velocities between -30 and -65~km~s$^{-1}$  distributed within the continuum source. 
A NLTE model, with density and temperature of the order $10^7$ cm$^{-3}$ and $10^4$ K,  reproduce both the observed H30$\alpha$ line profiles and  their  underlying continuum flux densities.  
Three of these sources are identified with Weigelt blobs D, C and B; estimating their proper motions,  we derive ejection times (in years) of 1952.6, 1957.1, and 1967.6, respectively, all of which are close to  periastron passage. 
Weaker H30$\alpha$ line emission is detected at higher positive and negative velocities,  extending in the direction of the Homunculus axis. 
The He30$\alpha$ recombination line is also detected with the same velocity of the narrow H30$\alpha$ line.   
Finally, the close resemblance of the H30$\alpha$ image with that of an emission line that was reported in the literature as HCO$^+$(4-3) led us to identify this line as H40$\delta$ instead, an identification that is further supported by modeling results. Future observations will enable to determine the proper motions of all the compact sources discovered in the new high-angular resolution data of $\eta$ Carinae. 
\end{abstract}


%%%%%%%%%%%%%%%%%%%%%%%%%%%%%%%%%%%%%%%%%%%%%%%%%%

%%%%%%%%%%%%%%%%% BODY OF PAPER %%%%%%%%%%%%%%%%%%

\section{Introduction}

\label{sec:Introduction}

The massive $(M\ga120~ M_\odot)$  and highly eccentric $(e \sim 0.9)$ binary system formed by $\eta$ Carinae (hereafter $\eta$ Car) and its unknown companion \citep{dam96,dah97,cor01,fal05} attracted the astronomers' attention after the ejection of a large amount of matter $(\ga 12~ M_\odot)$ in a dramatic eruption that started around 1840 and formed what it is now known as the Homunculus  Nebula, which expands  with a velocity of about 650 km s$^{-1}$ \citep{gav50,smi03}.
This event characterized $\eta$ Car as a probable Luminous Blue Variable (LBV) star: its brightness increased abruptly by about four magnitudes and   decreased steadily by nine magnitudes in 20 years, as dust formed and absorbed the visible light, re-emitting it in the infrared \citep{hum99}.

Another smaller eruption occurred in 1887, also characterized by a sudden increase in brightness that lasted for seven years \citepalias{hum99}, forming the Little Homunculus, detected inside the Homunculus  at a distance of about 2~arcsec from the binary system, and expanding with a velocity of about 200 km s$^{-1}$ \citep{ish03}.

A third, small and sudden increase in luminosity was observed around 1940, but instead of returning to its previous value, the luminosity continued to   slowly increase \citepalias{hum99}. This event coincided with the first detection of the spectroscopic events \citep{gav53}, in which
high excitation lines disappear from the spectrum for a short time during each binary orbit. 
Observations of $\eta$ Car obtained with the Atacama Large Millimeter Array (ALMA) in the continuum and various recombination lines led to the discovery of an unresolved cloud of high density ionized gas, whose dynamics correspond to matter ejected during the 1941 event and was therefore named "Baby Homunculus" \citep{abr14}.   

Besides those isolated episodes of mass ejection, $\eta$ Car  loses mass continuously  through a massive wind, with a mass loss rate $\dot{M} \sim 10^{-3}~ M_\odot~ \rm{yr}^{-1}$ and a velocity $v \sim 500$ km s$^{-1}$ \citep{gro12}. The companion star, although undetected, must  also be losing mass through a less massive but faster wind ($\dot{M} \sim 10^{-5}~ M_\odot~ \rm{yr}^{-1}$, $v \sim 3000$ km s$^{-1})$, in order to explain the observed X-ray emission arising from hot shocked material in the wind-wind collision zone \citep{cor01}.

The combination and superposition of all these processes complicate the interpretation of the images and spectra of $\eta$ Car and its surroundings, in addition to the variability with orbital phase and along the binary cycles.

Observations at UV and optical wavelengths show a series of broad emission lines, sometimes with P Cygni absorption profiles \citep{gul05,nie05,gul06,mar10}, as well as narrow emission lines, some of them from high excitation states \citep{dav97,mor98,gul09,meh10,gul16}. Those include the so-called Weigelt blobs, which are dense condensation of matter $(n_e \sim 10^7 {\rm cm}^{-3}, T_e \sim 10^4~ {\rm K})$, discovered in 1986 at distances of $0\farcs1 - 0\farcs3$ from the binary system \citep{wei86} that are moving away with velocities of several mas  yr$^{-1}$ \citep{hof88,dav97,smi04,dor04}.

Infrared images reveal a complex dust distribution around $\eta$ Car, implying that some of the UV and optical structures could be the result of selective absorption \citep{mor99,hon01,che05,art11,mor17}

The properties of the individual stars in the binary system are only inferred from indirect observations. The  mass of $\eta$ Car, according to the Eddington limit, should be larger than 120 M$_\odot$. Its mass loss rate is very  large and its photosphere cannot be observed directly, but a possible range of effective temperatures was calculated fitting its spectra to  models obtained from a non LTE-line blanketed code \citep{hil01,hil06}. Using the observed present bolometric luminosity of $5\times10^6$ L$_\odot$, inferred from the total emission of the Homunculus Nebula at a distance of 2.3 kpc, and the mass loss rate and terminal velocity mentioned before, they found an effective temperature $T_{\rm{eff}}\sim 30,000$ K, which gives a stellar radius of $R\sim 60$ R$_\odot$  and a rate of ionizing photons provided by the star of  $Q \sim 2 \times 10^{50}$ ph s$^{-1}$. 

 The spectra of the companion star is not detected, but its effective temperature was estimated from the number of ionizing photons with energies larger than 40 eV, necessary to account for the observed high ionization lines in the Weigelt blobs. Fitting radiative transfer models  to the high ionization  spectra of the blobs \citet{ver05} and \citet{meh10} showed that their electron densities should be  $\sim 10^6-10^7$ cm$^{-3}$ and the rate of Lyman ionizing photons from the secondary star  $Q \sim 10^{49}$ ph s$^{-1}$, corresponding to  effective temperatures of $\sim 37,000-40,000$~K.  Depending on the evolutionary state (main sequence, supergiant or Wolf-Rayet),  the mass of the companion  star should be $\sim 30-40$ M$_\odot$.

 Numerical simulations of the wind-wind collision in a region centered on $\eta$ Car and extending up to 100 times the semi-major axis of the binary orbit show a series of dense $(n_e \sim 10^7~ \rm{cm}^{-3})$ arc-like features formed by both the colliding winds and the winds ejected in the previous cycles \citep{par11,cle14,rus16} that are in good agreement with observed optical images  \citep{gul16}.

Radio frequencies are powerful tools for studying the matter distribution in the surroundings of $\eta$ Car, since they are not absorbed by dust.  Continuum and recombination line H91$\alpha$ and H106$\alpha$ images with arcsec resolution were obtained since 1992 with the Australian Compact Array (ATCA) at 3 and 6~cm, revealing a compact source, coincident with the central star and extended emission distributed around it with structures that changed with orbital phase \citep{whi94,dun95,dun97,dun03}. A non-resolved component centered at the star position was observed in the continuum and recombination lines during periastron passage, which evolved at apastron into an elongate $4\arcsec$ long ridge in the NE-SW direction, with a central velocity close to the star velocity and half power width (HPW) of 500 km s$^{-1}$. In addition, a compact component in the NW direction at a distance of $1\farcs2$ from the star was reported, with a velocity of -286~km~$^{-1}$ and  HPW of 250 km s$^{-1}$, and is related to the Little Homunculus \citep{smi05,teo08}. 

Single dish continuum observations at 1.3 and 2.9~mm with $30\arcsec$  and $50\arcsec$ resolution, respectively, showed a compact, non-resolved source, with orbital phase dependent flux density that increased with frequency \citep{cox95a}. H40$\alpha$, H30$\alpha$ and H29$\alpha$ recombination lines were also observed, presenting strong and narrow profiles, with velocities centered at -50 km~s$^{-1}$, revealing the existence of ionized gas in NLTE, with electron density of about $10^7$~cm$^{-3}$ \citep{cox95b}.

The surroundings of $\eta$ Car were further observed using ALMA in CO and HCN emission lines revealing a cool, disrupted equatorial torus located 4000 AU (corresponding to $1\farcs8$) from the star \citep{smi18}. \citet{bor19} also reported an extended structure northwest of $\eta$ Car seen in the continuum and in only one emission line, that they identified with HCO$^+$(4-3).

Previous continuum observations of the central region of $\eta$ Car with ALMA were done in Cycle 0, with frequencies and resolutions ranging from $100-700$~GHz and $3\arcsec-0\farcs4$, respectively,  and revealed  a spectrum characteristic of compact \ion{H}{ii} regions, with a turnover frequency around 300~GHz  \citepalias{abr14}. 
Strong narrow hydrogen recombination lines (H42$\alpha$, H40$\alpha$, H30$\alpha$, H28$\alpha$ and H21$\alpha$), with central velocities of $\sim -55$  km~s$^{-1}$ and HPW of $\sim 40$~km~s$^{-1}$ were also detected, showing departure from LTE condition.  Neither the continuum nor the recombination line images were resolved in these observations, even at the highest frequencies.

In this paper, we present new ALMA observations  with an order of magnitude higher resolution ($0\farcs065\times 0\farcs043$) than in previous observations, obtained in the continuum and in the recombination lines of H30$\alpha$  and He30$\alpha$. The new data reveal 16 individual unresolved components with different velocities, three of which are identified with Weigelt blobs B, C an D.
The paper is organised as follows: in Section \ref{sec:Observations}, we describe the observations; in Section \ref{sec:Results}, we present the results; in Section \ref{Discussion}, we discuss and analyze the main results; and, in Section \ref{sec:Conclusions},we outline the conclusions of this study.

In the paper, we assume that  $\eta$ Car\footnote{The ICRS coordinates of $\eta$ Car are: $\alpha(\rm J2000) =$ 10:45:3.5362, $\delta(\rm J2000) =$ -59:41:4.0534} is at a distance of 2300 pc \citep{dav97}, so that 0.1 arcsec corresponds to 224 AU. Over the 5.54 year orbital period, 100 km~s$^{-1}$ corresponds to a spatial motion of 110 AU or 50 mas on the plane of the sky. All the velocities are referred to the Local Standard of Rest (LSR); in the direction of $\eta$ Car, the difference between LSR and heliocentric velocity is $-11.6$ km s$^{-1}$; in the LSR, the velocity of $\eta$ Car is $-19.7$ km s$^{-1}$ \citep{smi04}.
%%%%%%%%%%%%%%%%%%%%%%%%%%%%%%%%%%%%%%%%%%%%%%%%%%%%%%%%%%%%%%%%%%%%%%%%%%%%%%%
\section{Observations}
\label{sec:Observations}

The observations were performed in  November 20, 2017 with ALMA band 6 receivers,  using 43 antennas of the 12~m array, as part of Cycle 5 program 2017.1.00725.S. The closest baseline was 97.1~m, constraining the maximum recovery scale to $0\farcs8$, while the largest baseline was 8457.6~m, resulting in an angular resolution of 50 mas. The total observation time was 30 minutes, of which 13 minutes were on the source. The data is divided into 4 spectral windows, the main one centered at the frequency of the H30$\alpha$ recombination line (231.900~GHz) and the others, focused on the continuum emission, centered at 230.519, 218.018, and 215.518~GHz. The line spectral window has 1.875~GHz bandwidth, with channel width resolution of 976~kHz, producing a datacube of 1920 channels. The continuum bands have a bandwidth of 2~GHz divided into 128 channels. The pointing, focusing, amplitude, and flux calibration were performed through observations of the quasar J0904-5735, while quasar J1032-5917 was the phase calibrator. Data were processed using the  CASA 5.4.0 version.

The continuum image was obtained using the continuum bands plus the channels of the main spectral window where the line is not present. The spectral cube was obtained  for the spectral window of the recombination line after continuum subtraction. The convolution method used the Hogbom algorithm with Briggs weighting. The synthesized beam has a major axis of 65~mas and minor axis of 43~mas for both continuum and emission line images, and for both cases, the images have a cell size of 0.1~mas and a total size of $2048 \times 2048$ pixels.  In velocity, the cube range goes from $-1187$ to 1235~km~s$^{-1}$, with a velocity resolution of 1.262~km~s$^{-1}$. 

%%%%%%%%%%%%%%%%%%%%%%%%%%%%%%%%%%%%%%%%%%%%%%%%%%%%%%%%%%%%%%%%%%%%%%%%%%%%%%%
\section{Results}
\label{sec:Results}
%%%%%%%%%%%%%%%%%%%%%%%%%%%%%%%%%%%%%%%%%%%%%%%%%%%%%%%%%%%%%%%%%%%%%%%%%%%%%%%
\subsection{Continuum emission}
\label{subsec:Continuum}

\begin{figure}
\begin{center}
\includegraphics[width=\columnwidth]{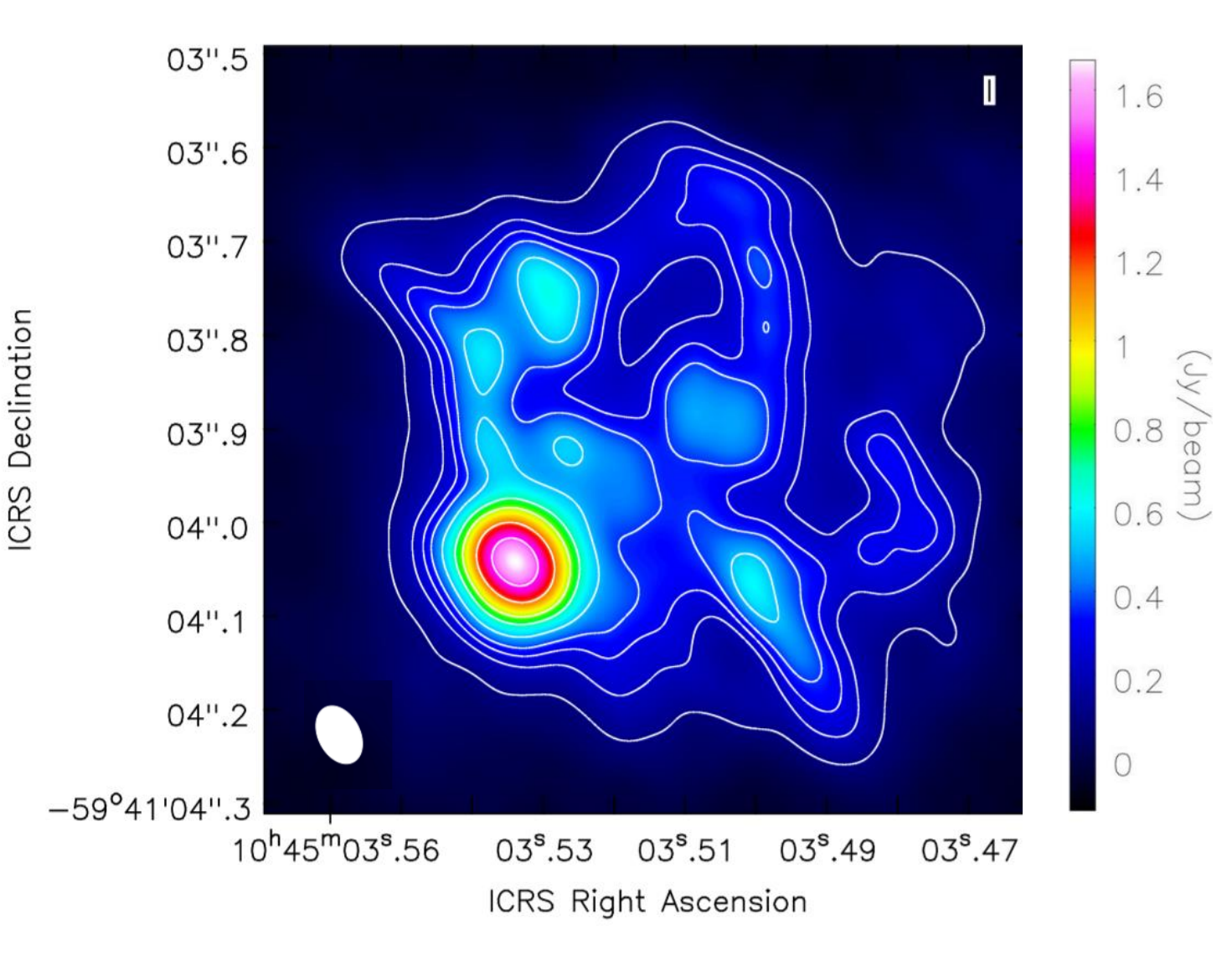}
\caption{230 GHz continuum image (raster and contours) of $\eta$ Car. Contours are 0.1, 0.17, 0.22, 0.27, 0.36, 0.5, 0.6, 0.8 and 1.0 Jy beam$^{-1}$. The scale of the continuum flux
is displayed on the right axis of the panel. The $65\times43$ mas synthesized beam (angle $24\fdg5$) is shown in white at the bottom.} 
\label{fig:Fig_1.pdf}
\end{center}
\end{figure}

Figure \ref{fig:Fig_1.pdf}  presents the 230 GHz continuum image of the $0\farcs8 \times 0\farcs8$ (1800 AU) central region of $\eta$ Car. It shows a strong compact component, coincident in position with the binary system, and emission extending north-west of the compact source. 

The integrated 230 GHz continuum flux density of the observed region is $28.3 \pm 0.1$ Jy.
The compact source HPW  is  $0\farcs13 \times 0\farcs11$, in the directions of the major and minor axis of the beam  ellipse. 
Its flux density  is $4.5 \pm 0.1$ Jy and  although its size is larger than the half power beam width (HPBW),  no structure is observed.
The extended emission covers about $0\farcs6$ and appears structured with regions of enhanced emission. 

Previous ALMA continuum imaging results at 346 GHz, published by \citet{bor19}, already showed a similar structure with a point-like source and  an extended emission in the north-west direction (which they labeled "Peanut"), although these data were taken at a lower angular resolution ($0\farcs17 \times 0\farcs13$) and signal-to-noise ratio.

%%%%%%%%%%%%%%%%%%%%%%%%%%%%%%%%%%%%%%%%%%%%%%%%%%%%%%%%%%%%%%%%%%%%%%%%%%%%%%
%%%%%%%%%%%%%%%%%%%%%%%%%%%%%%%%%%%%%%%%%%%%%%%%%%%%%%%%%%%%%%%%%%%%%%%%%%%%%%
\begin{figure*}
\begin{center}
\includegraphics[width= 14cm]{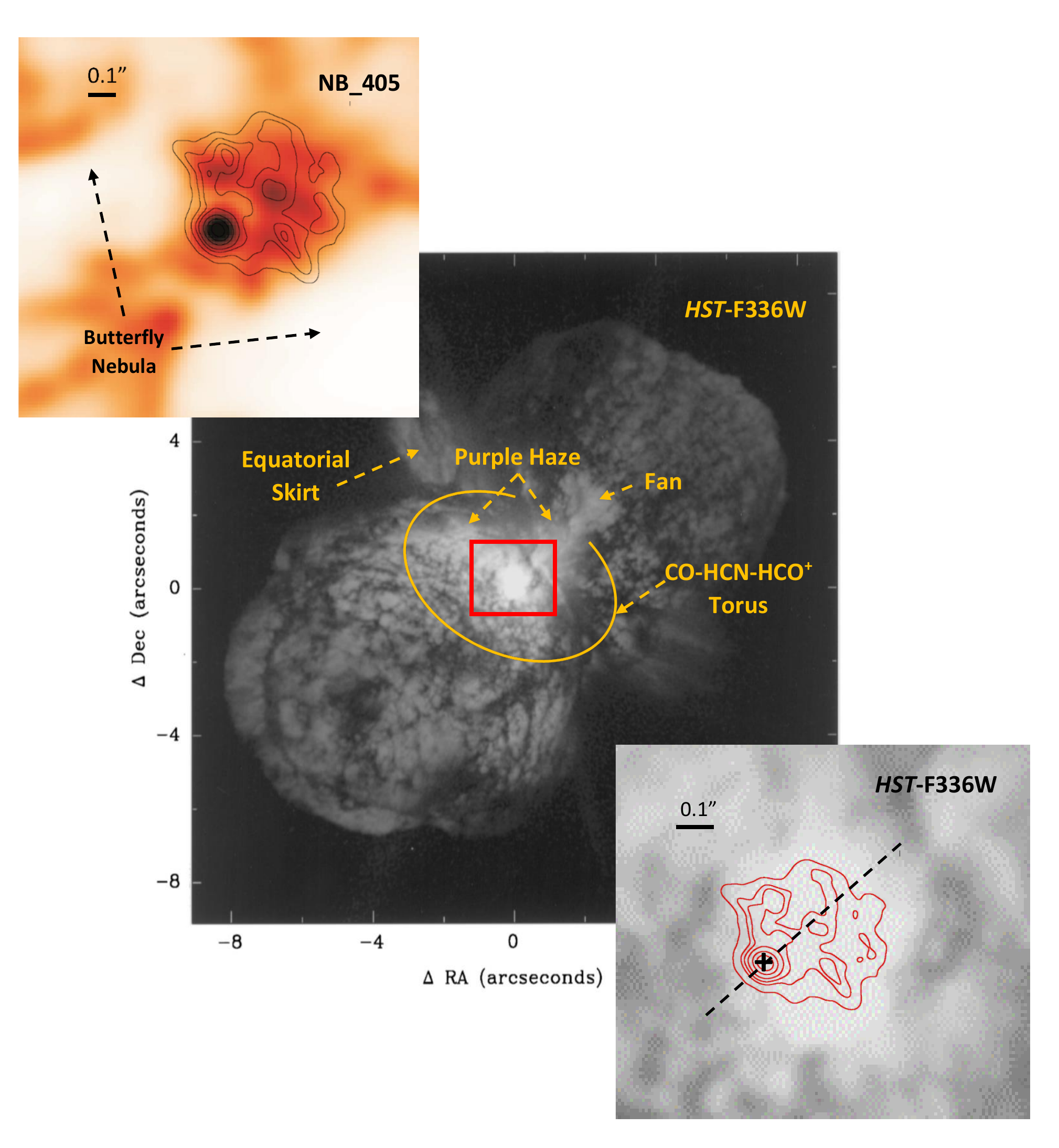}
\caption{Central image: $HST$ Wide Field Planetary Camera image of the Homunculus  in the F336W filter \citep{mor98}; the red rectangle delimits the size and position of the images in the lower right corner; the main features of the nebula are identified. Bottom right image: 230 GHz continuum contours superimposed to the $HST$ image; the black cross represents the position of $\eta$ Car and the black broken line shows the direction of the position-velocity PV diagram presented in Figure \ref{fig:Fig_3.pdf}. Upper left image: 230 GHz continuum contours superimposed to the NB\_405 narrow filter image, which includes the Br$\alpha$ line, obtained with the VLT Adaptive Optics system NACO \citep{che05}}
\label{fig:Fig_2.pdf}
\end{center}
\end{figure*}
%%%%%%%%%%%%%%%%%%%%%%%%%%%%%%%%%%%%%%%%%%%%%%%%%%%%%%%%%%%%%%%%%%%%%%%%%%%%%%
\begin{figure*}
\begin{center}
\includegraphics[width= 14cm]{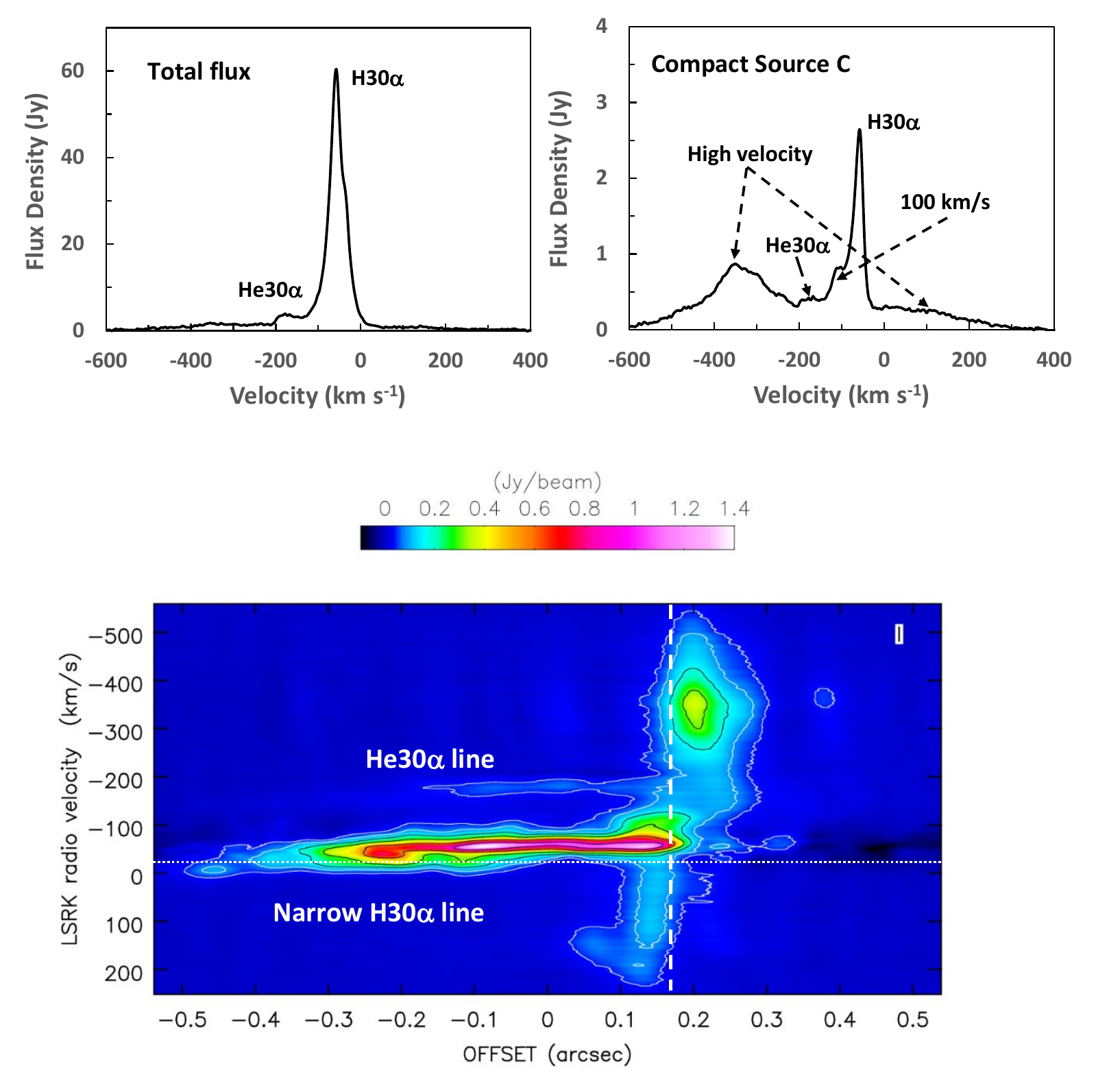}
\caption{Top left: spectrum of the H30$\alpha$ line, integrated over the whole continuum emitting region. Top right: spectrum  integrated over a circle of $0\farcs09$ radius, centered on the continuum compact source. Bottom: PV diagram of the region, collapsed along the line shown in the bottom right image of Figure \ref{fig:Fig_2.pdf}, in the direction of the Homunculus axis ($42\degr$), with a width of 10 mas, which includes the position of the compact continuum source, shown as a vertical broken line.
 The horizontal dotted line at $-19.7$ km s$^{-1}$ represents the LSR velocity of $\eta$ Car.} Contours are: 0.03, 0.06 (white), 0.12, 0.22, 0.4 and 0.7 (black) of 1.403 Jy beam$^{-1}$
\label{fig:Fig_3.pdf}
\end{center}
\end{figure*}
%%%%%%%%%%%%%%%%%%%%%%%%%%%%%%%%%%%%%%%%%%%%%%%%%%%%%%%%%%%%%%%%%%%%%%%%%%%%%
\begin{figure*}
\begin{center}
\includegraphics[width= 14cm]{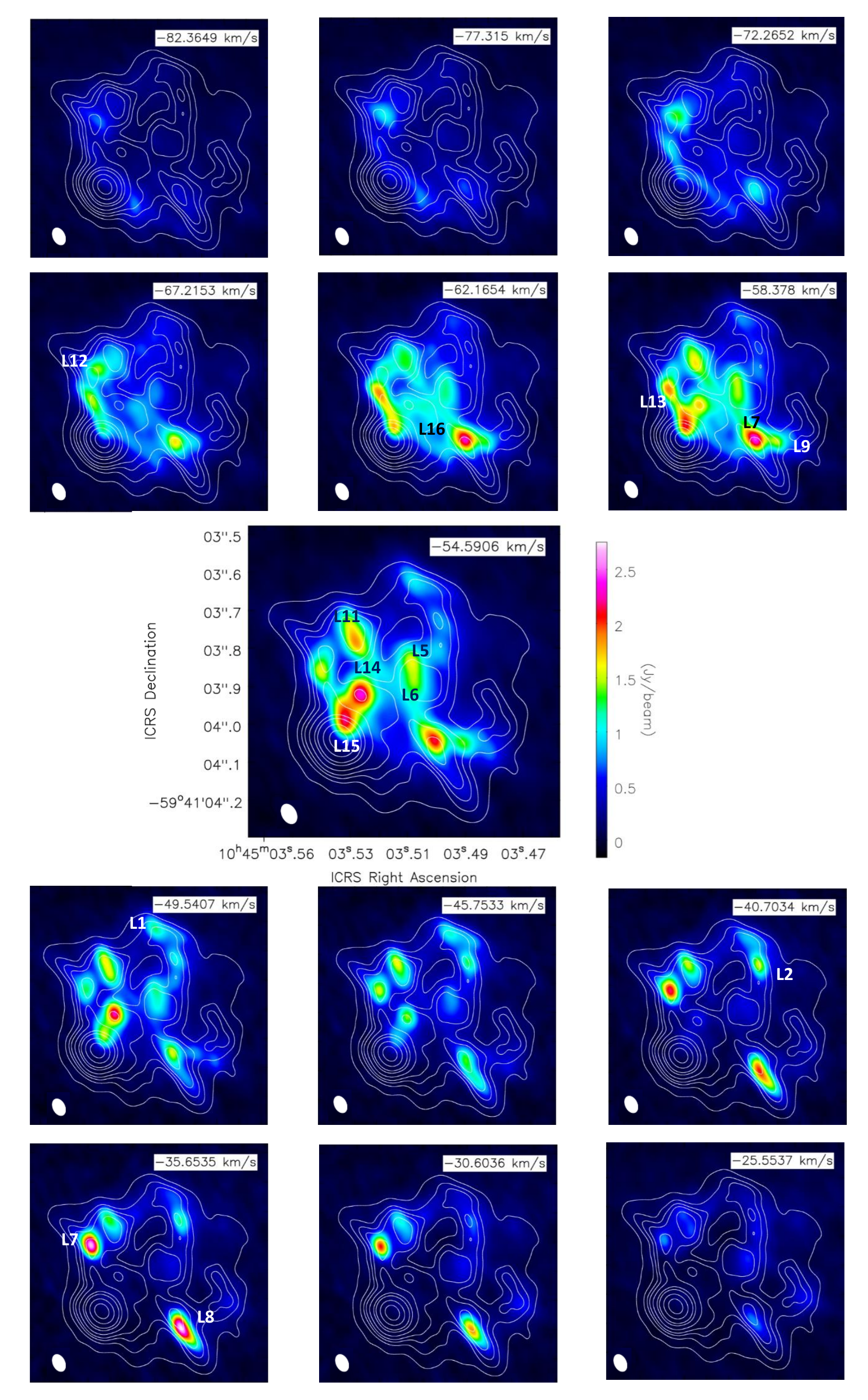}
\caption{Velocity-integrated intensity maps of channels with width of 1.262 km~s$^{-1}$, showing the H30$\alpha$ narrow emission line. The channels are separated in intervals of about 5 km~s$^{-1}$, between -82 and -26 km~s$^{-1}$.   All the maps have the same intensity scale (shown on the right axis of the middle panel). The center velocities of the H30$\alpha$ map are indicated in the upper right corners of each panel.  The strongest of the compact sources presented in Table \ref{tab:Table_1} (L1, L2, L5 - L9, L11 - L16) are also identified in the images.  The H30$\alpha$ maps are superimposed on the 230 GHz continuum map shown as contours (using the same levels as in Fig. \ref{fig:Fig_1.pdf})}. 
\label{fig:Fig_4.pdf}
\end{center}
\end{figure*}
%%%%%%%%%%%%%%%%%%%%%%%%%%%%%%%%%%%%%%%%%%%%%%%%%%%%%%%%%%%%%%%%%%%%%%%%%%%%%%
\begin{figure}
\begin{center}
\includegraphics[width=\columnwidth]{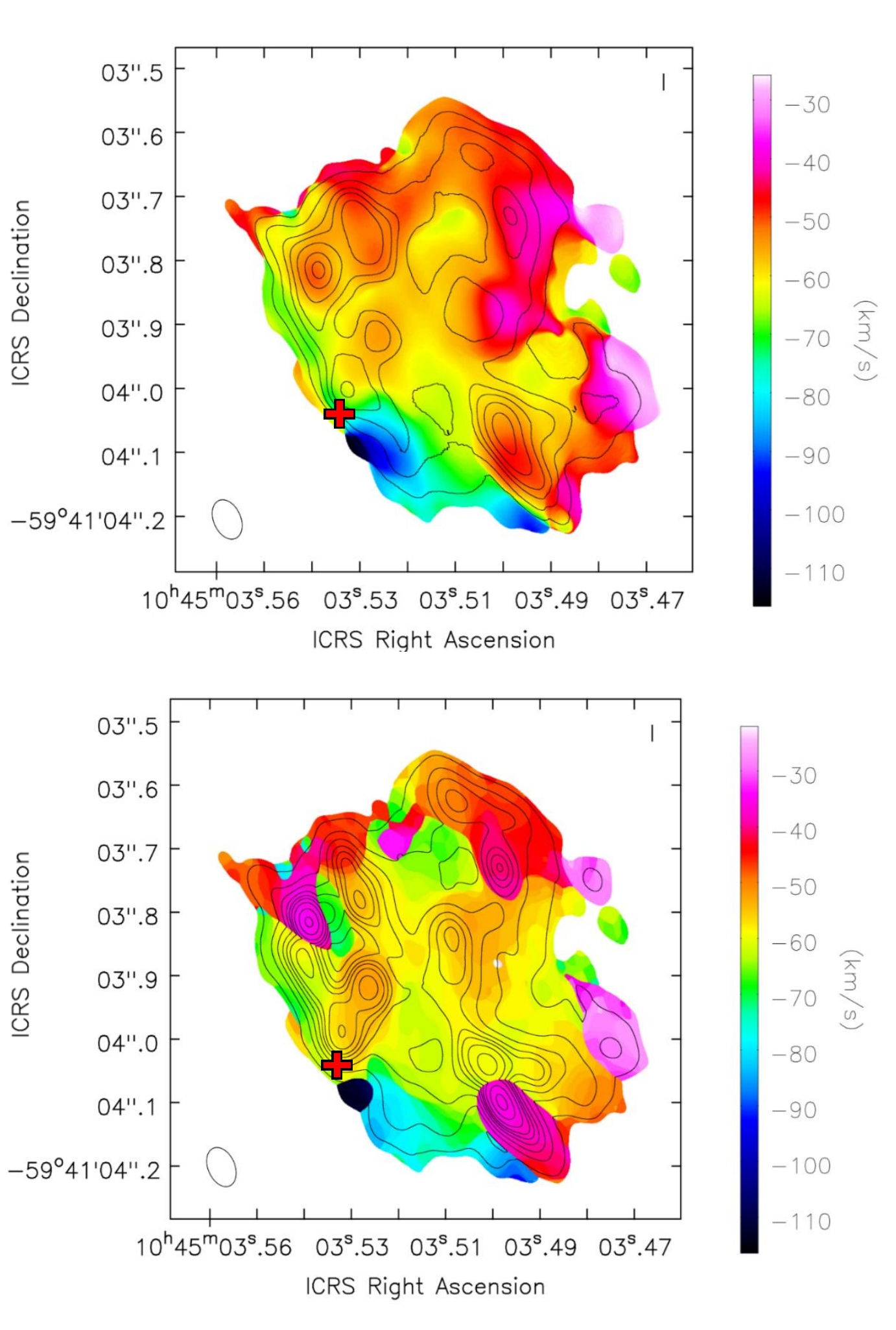}
\caption{Top image: The zero momentum image of the H30$\alpha$ recombination line (shown as contours) between $-116$ and $-25$ km s$^{-1}$ superimposed on the raster image of the first momentum map.  Contours represent 0.1, 0.2, 0.35, 0.45, 0.6, 0.8 and 0.95 of 89.5 Jy beam$^{-1}$ km s$^{-1}$. The red cross represents the position of $\eta$ Car. Bottom image: The maximum flux density of the H30$\alpha$ line (contours) superimposed on the raster image of the velocity of the line maximum; contours are 0.1, 0.2, 0.3, 0.4, 0.48, 0.55, 0.65, 0.8 and 0.95 of 2.76 Jy beam$^{-1}$. }
\label{fig:Fig_5.pdf}
\end{center}
\end{figure}
%%%%%%%%%%%%%%%%%%%%%%%%%%%%%%%%%%%%%%%%%%%%%%%%%%%%%%%%%%%%%%%%%%%%%%%%%%%%%%%
\begin{figure*}
\begin{center}
\includegraphics[width=16cm]{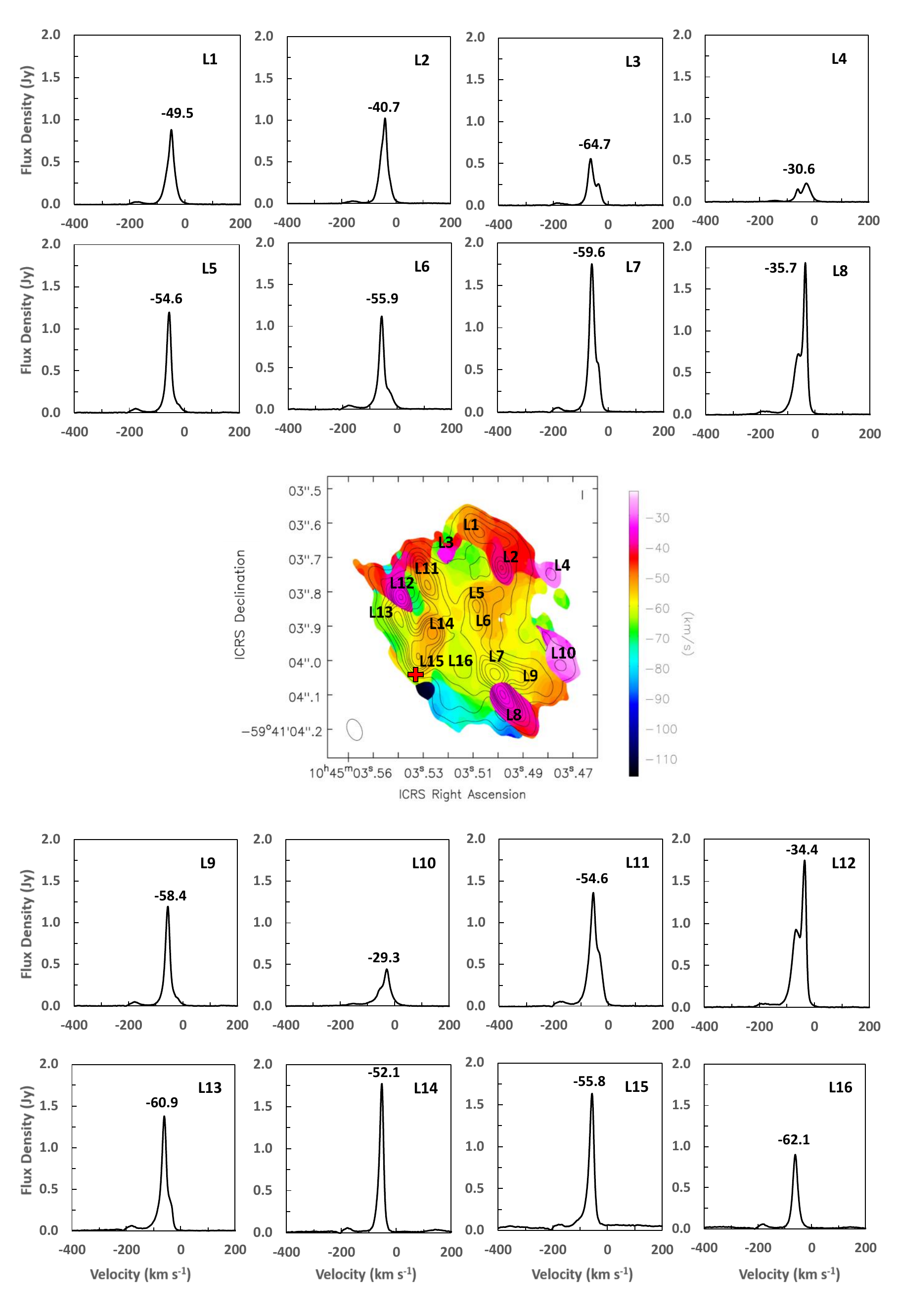}
\caption{Individual spectra of the sixteen regions identified in the first moment image of H30$\alpha$ (L1-L16), integrated over a circle of 66~mas diameter. The velocities are identified by the color scale shown on the right of the central panel.   The numbers on top of the lines represent the peak velocity in km s$^{-1}$. The red cross in the image represents the position of the compact continuum source.}
\label{fig:Fig_6.pdf}
\end{center}
\end{figure*}
%%%%%%%%%%%%%%%%%%%%%%%%%%%%%%%%%%%%%%%%%%%%%%%%%%%%%%%%%%%%%%%%%%%%%%%%%%%%%
\begin{figure*}
\begin{center}
\includegraphics[width= 13cm]{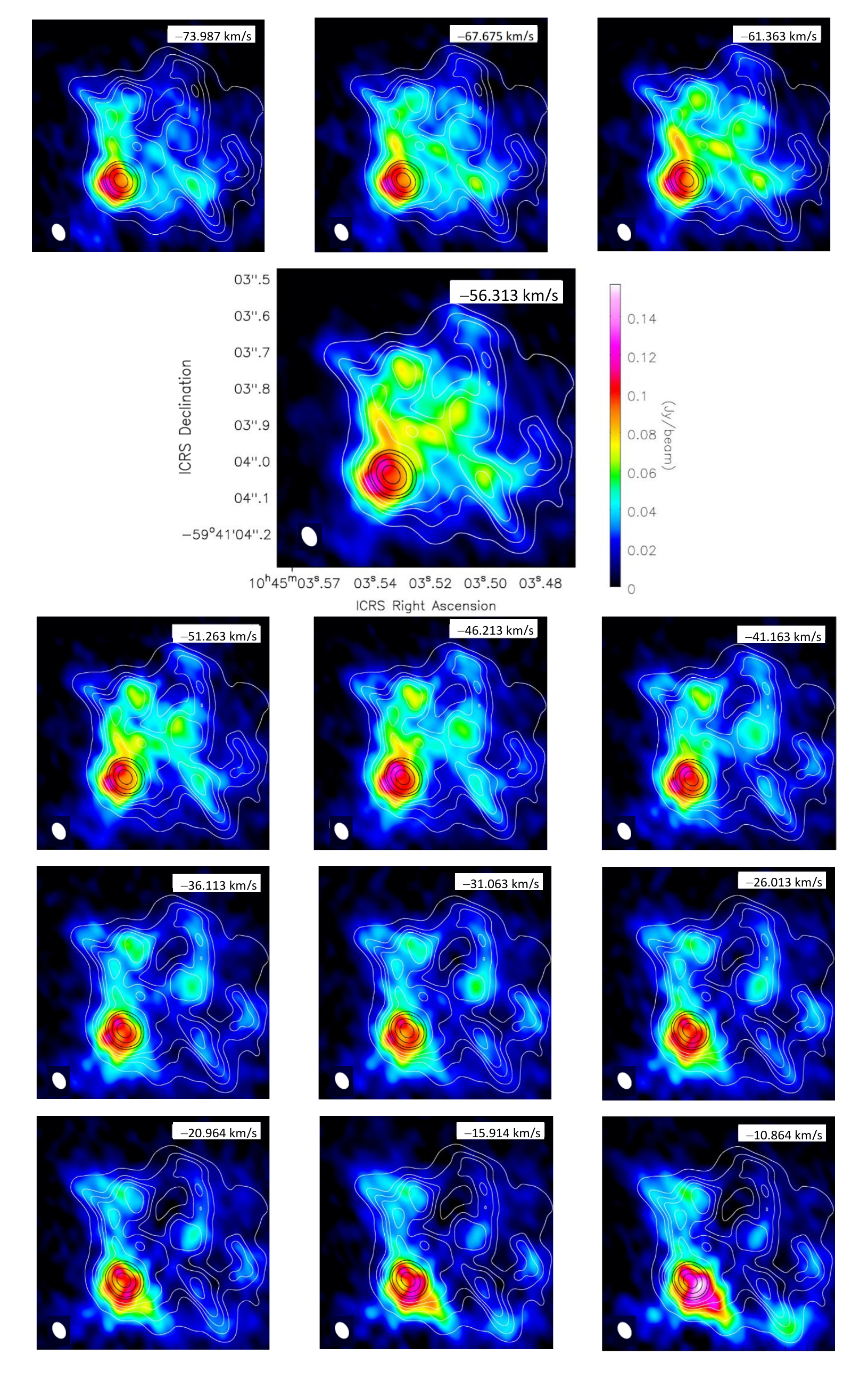}
\caption{Velocity-integrated intensity maps of the He30$\alpha$ emission line, integrated in bins of 1.262 km~s$^{-1}$. All the maps have the same intensity scale (shown on the right axis of the middle panel). The center velocities, indicated in the upper right corners of each panels, correspond to frequencies centered on the H30$\alpha$ line. The center velocities, indicated in the upper right corners of each panel, correspond to frequencies centered on the He30alpha line. The He30$\alpha$ maps are superimposed on the 230 GHz continuum map shown as contours (using the same levels as in Fig. \ref{fig:Fig_1.pdf}). The synthesized beam is shown in the  bottom left corner.}   
\label{fig:Fig_7.pdf}
\end{center}
\end{figure*}
%%%%%%%%%%%%%%%%%%%%%%%%%%%%%%%%%%%%%%%%%%%%%%%%%%%%%%%%%%%%%%%%%%%%%%%%%%%%%%
\begin{figure*}
\begin{center}
\includegraphics[width= 13cm]{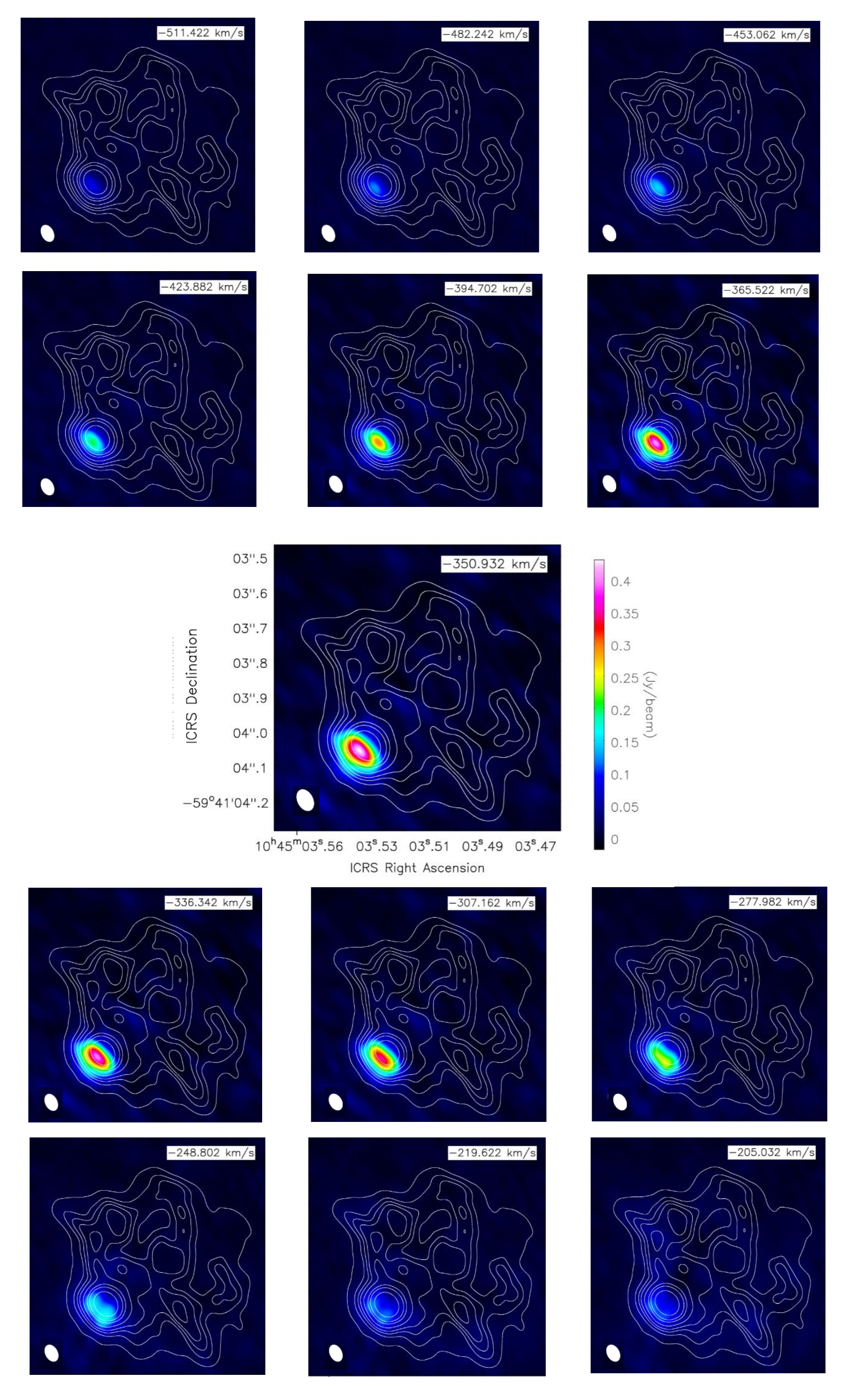}
\caption{Velocity-integrated intensity maps of the high negative velocity  H30$\alpha$  emission line, integrated in bins of 14.6 km~s$^{-1}$. All the maps have the same intensity scale (shown on the right axis of the middle panel). The center velocities are indicated in the upper right corners of each panel.  The H30$\alpha$ maps are superimposed on the 230 GHz continuum map shown as contours (using the same levels as in Fig. \ref{fig:Fig_1.pdf}). The synthesized beam is shown in the bottom left corner.}
\label{fig:Fig_8.pdf}
\end{center}
\end{figure*}
%%%%%%%%%%%%%%%%%%%%%%%%%%%%%%%%%%%%%%%%%%%%%%%%%%%%%%%%%%%%%%%%%%%%%%%%%%%%%
\begin{figure*}
\begin{center}
\includegraphics[width= 13cm]{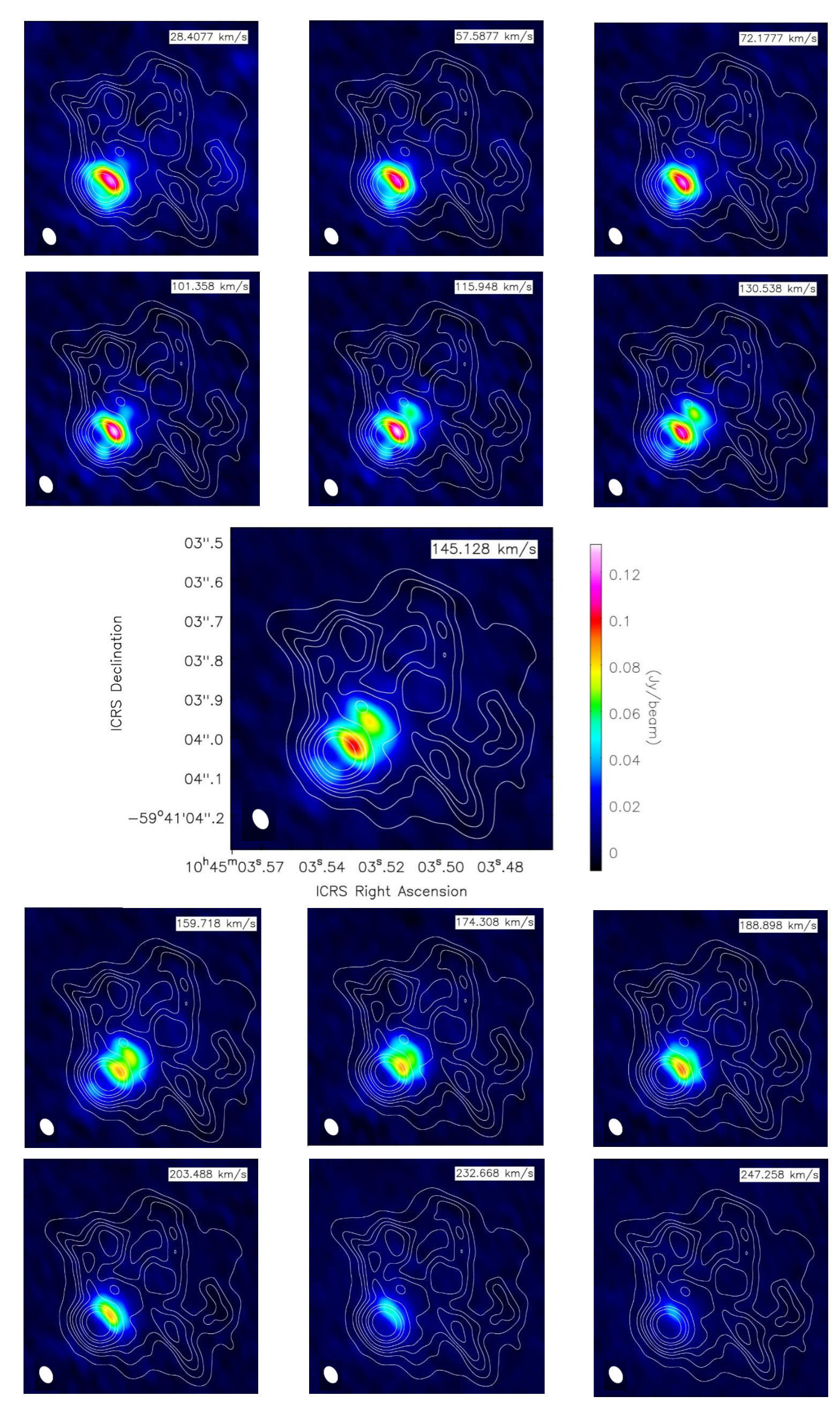}
\caption{Velocity-integrated intensity maps of the high positive velocity  H30$\alpha$  emission line, integrated in bins of 14.6 km~s$^{-1}$. All the maps have the same intensity scale (shown on the right axis of the middle panel). The center velocities are indicated in the upper right corners of each panels.  The H30$\alpha$ maps are superimposed on the 230 GHz continuum map shown as contours (using the same levels as in Fig. \ref{fig:Fig_1.pdf}). The synthesized beam is shown in the  bottom left corner.}
\label{fig:Fig_9.pdf}
\end{center}
\end{figure*}
%%%%%%%%%%%%%%%%%%%%%%%%%%%%%%%%%%%%%%%%%%%%%%%%%%%%%%%%%%%%%%%%%%%%%%%%%%%%%%
\begin{figure}
\begin{center}
\includegraphics[width=\columnwidth]{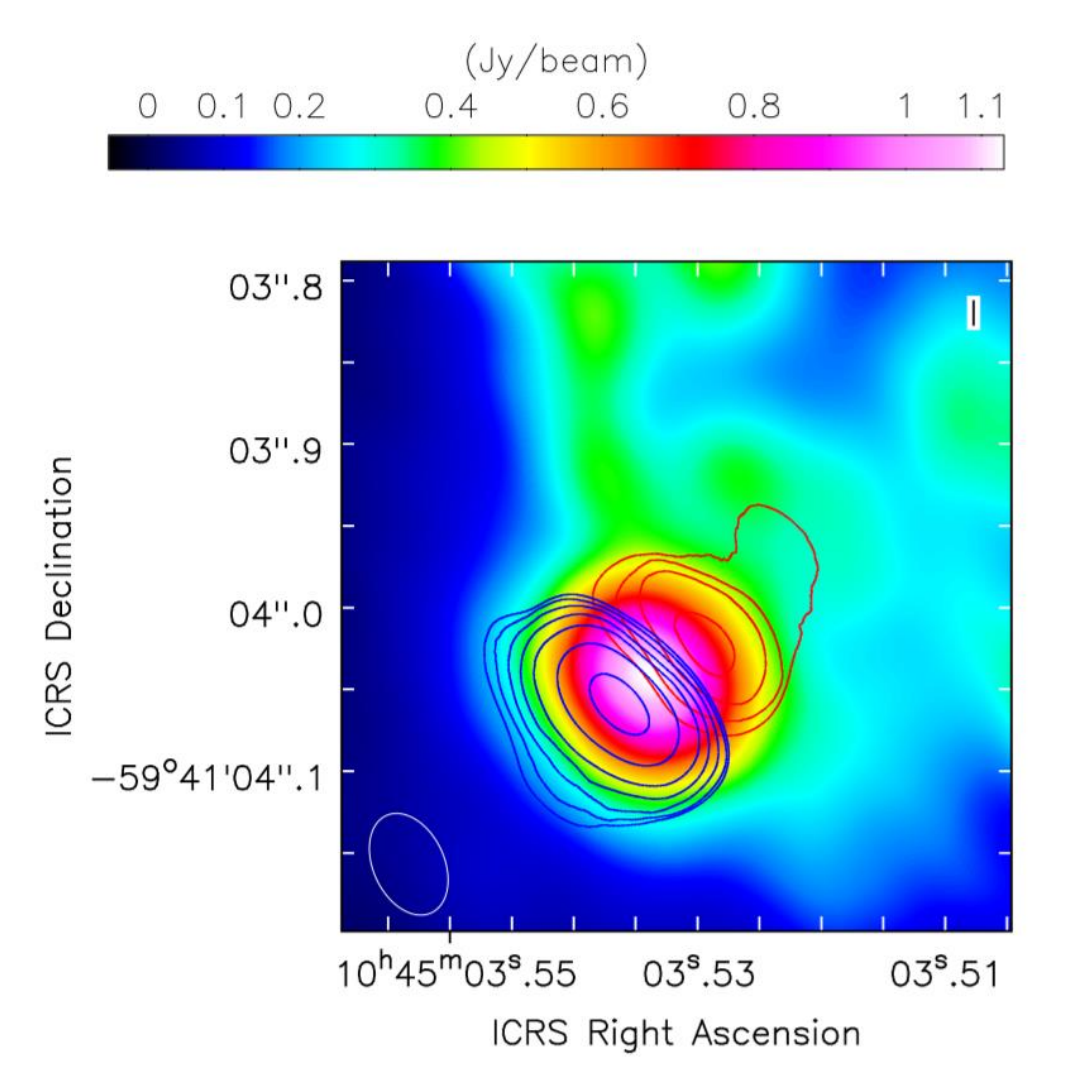}
\caption{Contour map of the positive and negative high velocity H30$\alpha$ emission, integrated over the line profile between velocities 22 and 250 km~s$^{-1}$ and -590 and -252 km~s$^{-1}$, respectively, superimposed on the continuum raster image. Blue  contours represent negative velocities,  red contours positive velocities. The contours are: : 0.05, 0.12, 0.2, 0.3, 0.4, 0.6 and 0.9 of 64.6 Jy~beam$^{-1}$ km~s$^{-1}$. }
\label{fig:Fig_10.pdf}
\end{center}
\end{figure}
%%%%%%%%%%%%%%%%%%%%%%%%%%%%%%%%%%%%%%%%%%%%%%%%%%%%%%%%%%%%%%%%%%%%%%%%%%%%%%%
\begin{figure*}
\begin{center}
\includegraphics[width= 13cm]{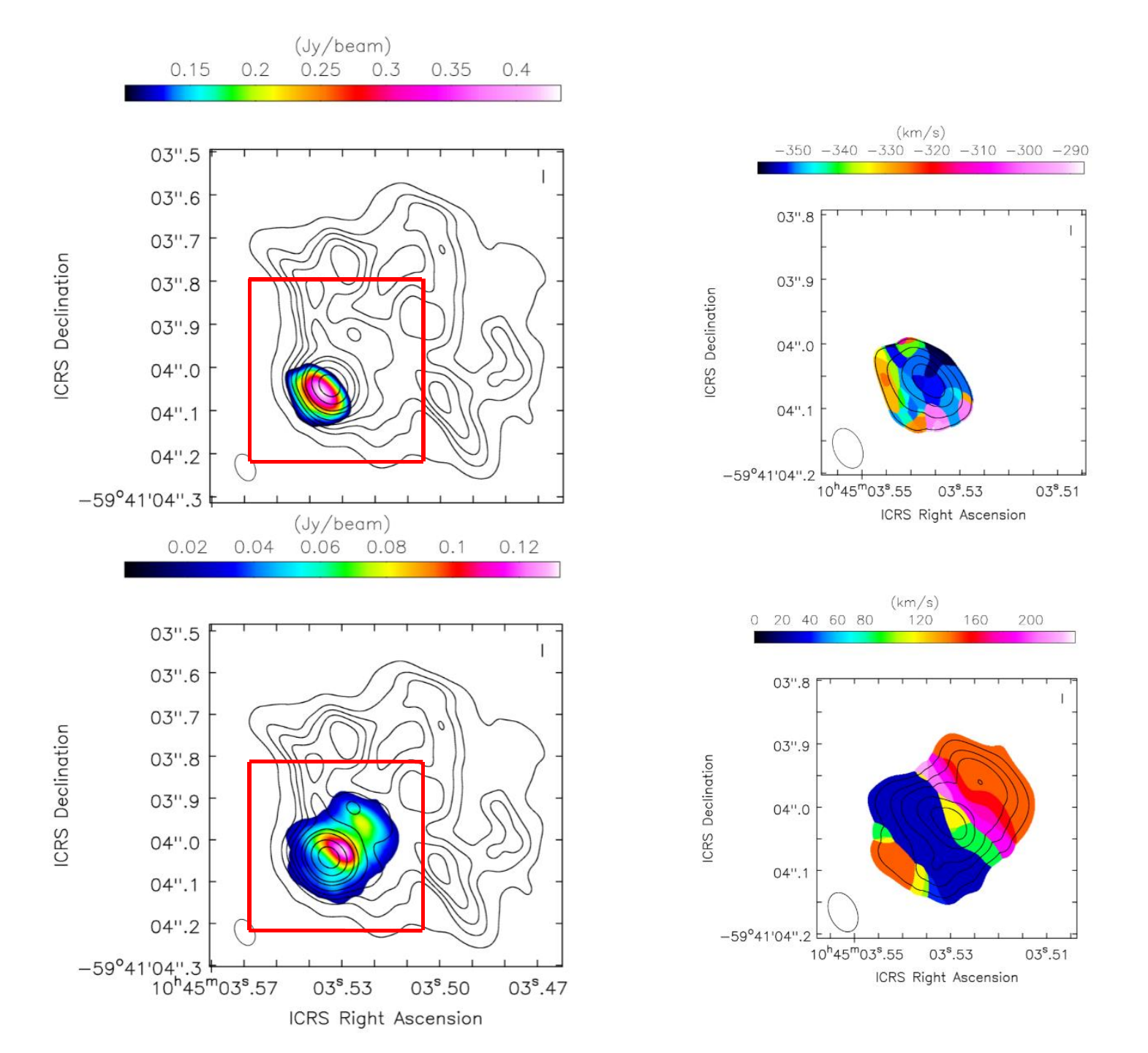}
\caption{Top left: raster image of the high negative velocity line emission zero momentum, integrated between $-286$ and $-357$ km s$^{-1}$, superimposed to the continuum contour map. Bottom left: the same as top left, but for  high positive velocities integrated between  28 and 232 km s$^{-1}$ .Continuum map contour levels are the same as in Figure \ref{fig:Fig_1.pdf}. Red squares represent the inserts shown to the figure right. Top right: contour map of the negative high velocity zero momentum, superimposed to the velocity weighted (first momentum) raster image. Contours are 0.25, 0.4, 0.6 and 0.8 of 0.435 Jy beam$^{-1}$ km s$^{-1}$. Bottom right: the same as top right for high positive velocities. Contours are: 0.2, 0.3, 0.4, 0.6 and 0.8 of 0.132 Jy beam$^{-1}$ km s$^{-1}$}
\label{fig:Fig_11.pdf}
\end{center}
\end{figure*}
%%%%%%%%%%%%%%%%%%%%%%%%%%%%%%%%%%%%%%%%%%%%%%%%%%%%%%%%%%%%%%%%%%%%%%%%%%%%%
\begin{figure*}
\begin{center}
\includegraphics[width=12cm]{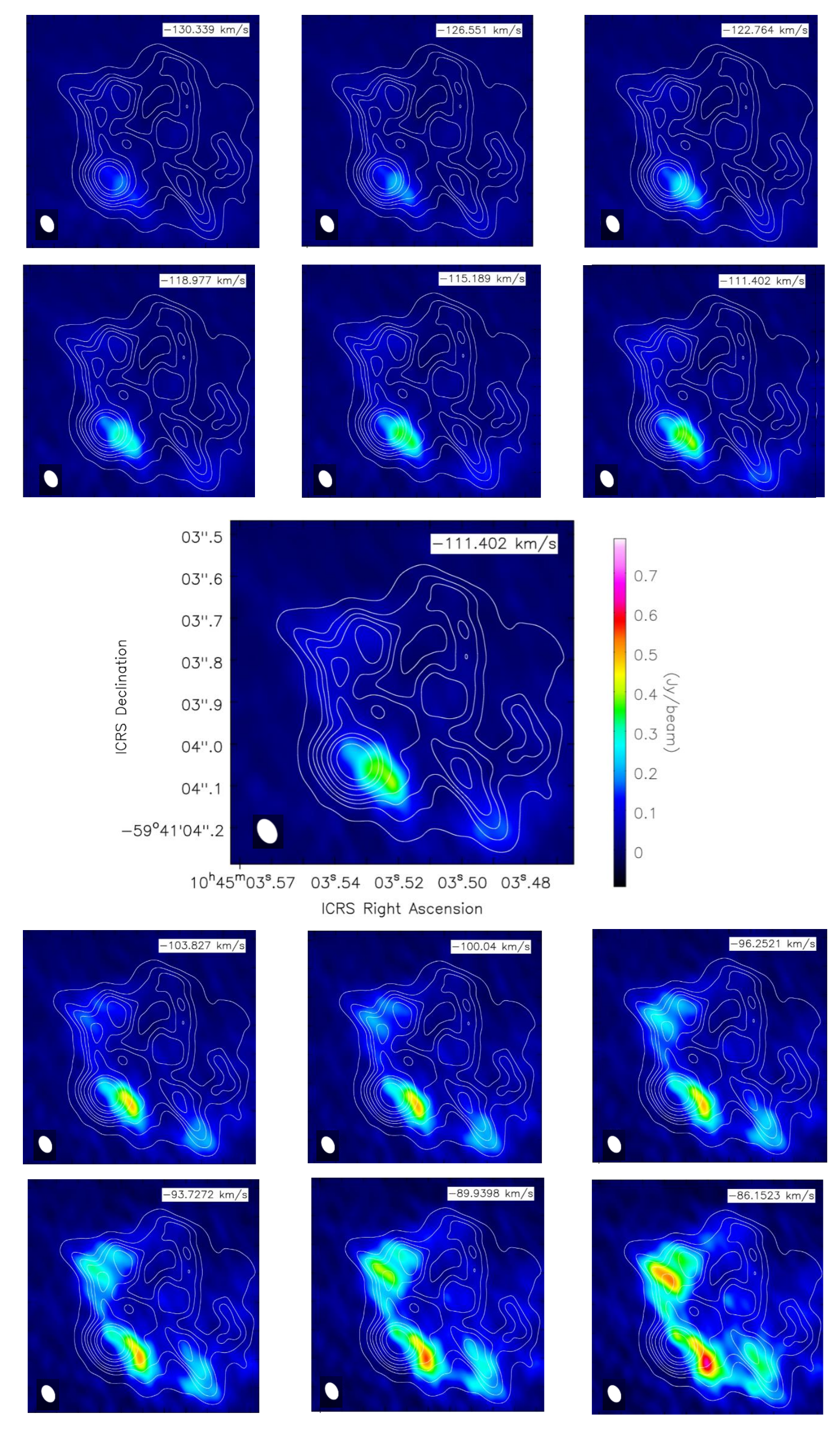}
\caption{Velocity-integrated intensity maps of the -100 km~s$^{-1}$ component of the  H30$\alpha$  emission line, integrated in bins of 1.262 km~s$^{-1}$. All the maps have the same intensity scale (shown on the right axis of the middle panel). The center velocities are indicated in the upper right corners of each panels.  The H30$\alpha$ maps are superimposed on the 230 GHz continuum map shown as contours (using the same levels as in Fig. \ref{fig:Fig_1.pdf}). The synthesized beam is shown at the  bottom left corner.}
\label{fig:Fig_12.pdf}
\end{center}
\end{figure*}
%%%%%%%%%%%%%%%%%%%%%%%%%%%%%%%%%%%%%%%%%%%%%%%%%%%%%%%%%%%%%%%%%%%%%%%%%%%%%%%
\begin{figure}
\begin{center}
\includegraphics[width=6cm]{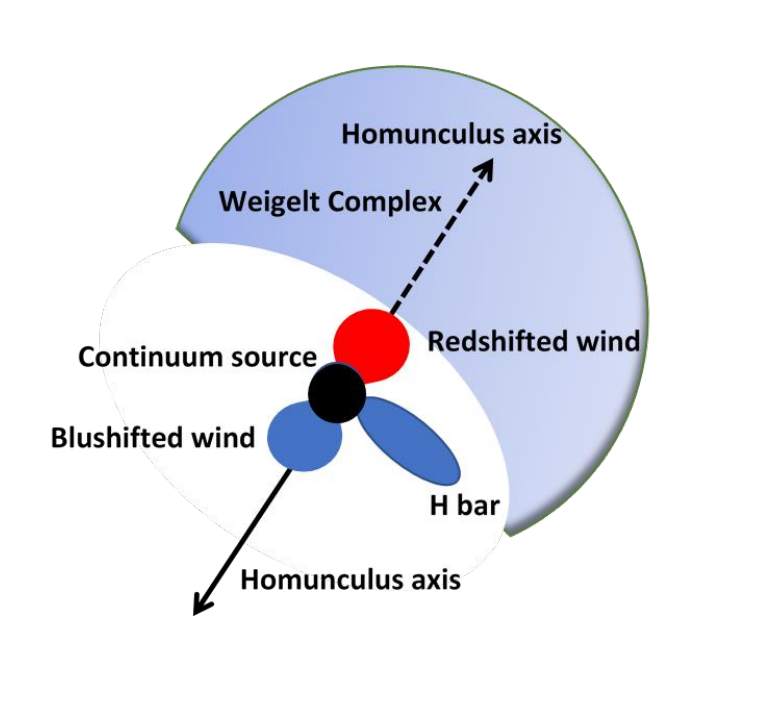}
\caption{Schematic showing the different features detected in the continuum and H30$\alpha$ images of $\eta$ Car}
\label{fig:Fig_13.pdf}
\end{center}
\end{figure}
%%%%%%%%%%%%%%%%%%%%%%%%%%%%%%%%%%%%%%%%%%%%%%%%%%%%%%%%%%%%%%%%%%%%%%%%%%%%%%%
\begin{figure}
\begin{center}
\includegraphics[width=\columnwidth]{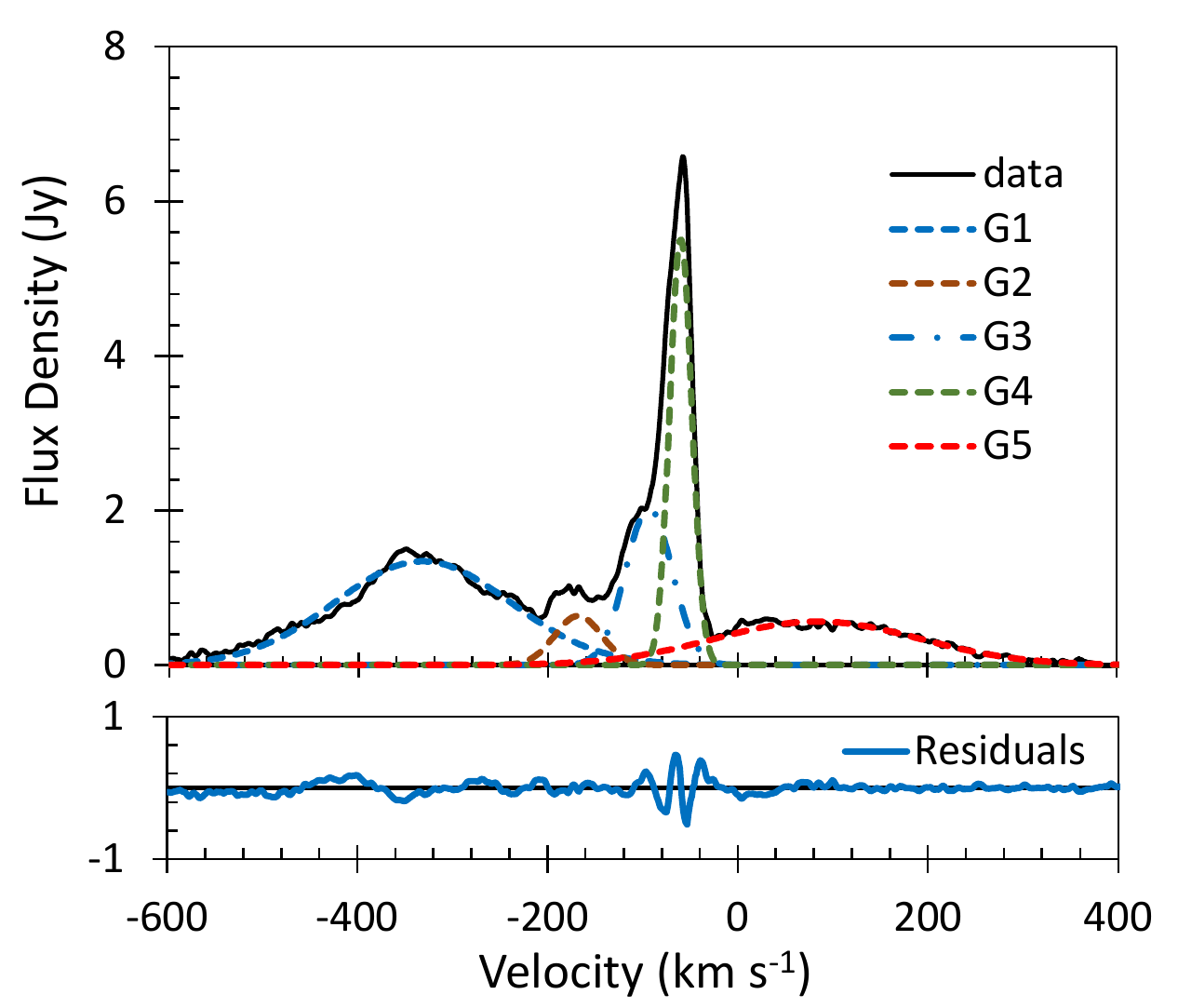}
\caption{H30alpha spectrum of the central source (integrated over a 90 mas radius region) and the five individual Gaussian components fitting the line (see text).}
\label{fig:Fig_14.pdf}
\end{center}
\end{figure}
%%%%%%%%%%%%%%%%%%%%%%%%%%%%%%%%%%%%%%%%%%%%%%%%%%%%%%%%%%%%%%%%%%%%%%%%%%%%%%%
\begin{figure}
\begin{center}
\includegraphics[width=\columnwidth]{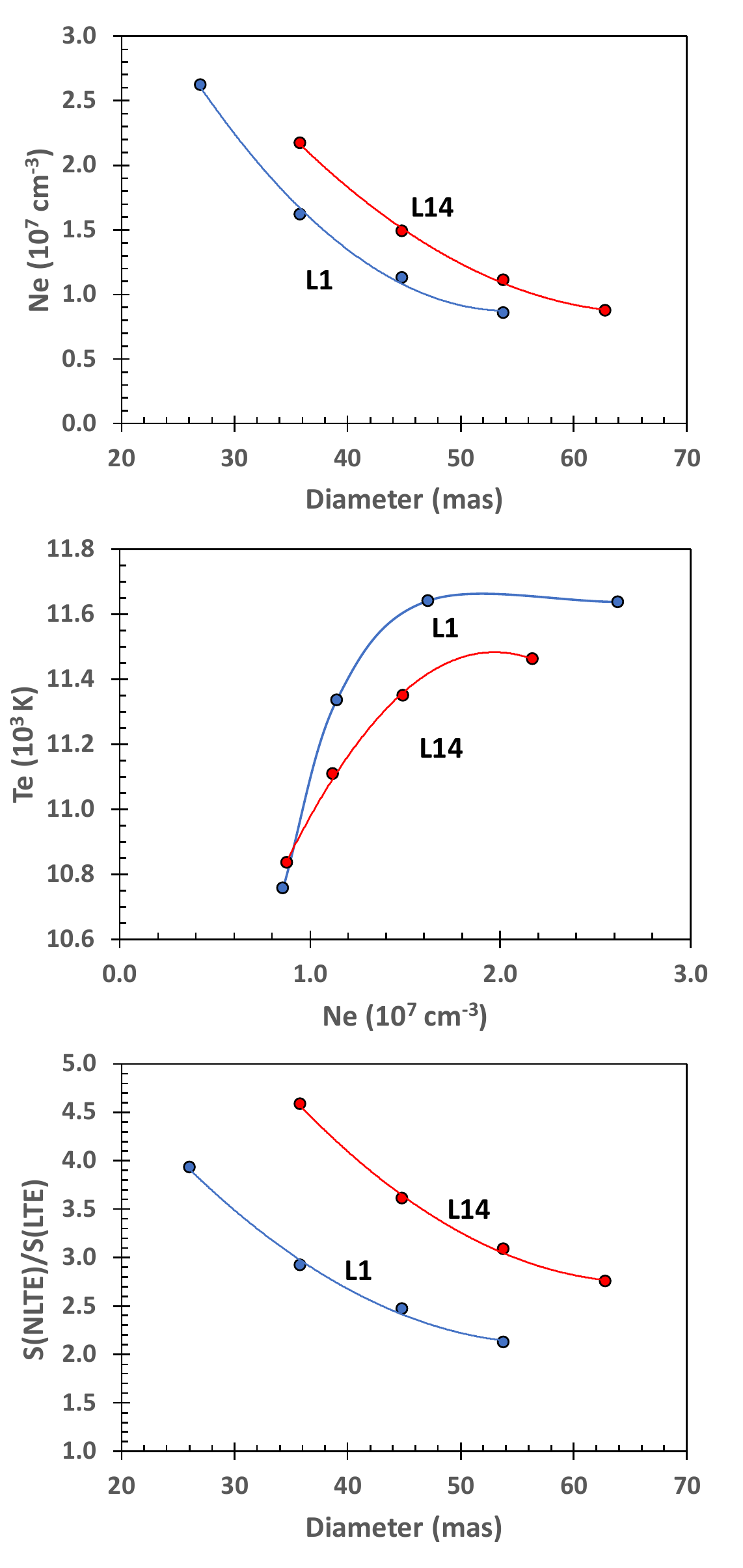}
\caption{Physical parameters of homogeneous spheres that fit the continuum flux density and maximum in the H30$\alpha$ line profile of blobs L1 and L14. Top: electron density $N_{\rm e}$ vs. source diameter; middle: electron temperature $T_{\rm e}$ vs. electron density $N_{\rm e}$; bottom: ratio of NLTE to LTE maximum flux density}
\label{fig:Fig_15.pdf}
\end{center}
\end{figure} 
%%%%%%%%%%%%%%%%%%%%%%%%%%%%%%%%%%%%%%%%%%%%%%%%%%%%%%%%%%%%%%%%%%%%%%%%%%%%%%%
\begin{figure}
\begin{center}
\includegraphics[width=\columnwidth]{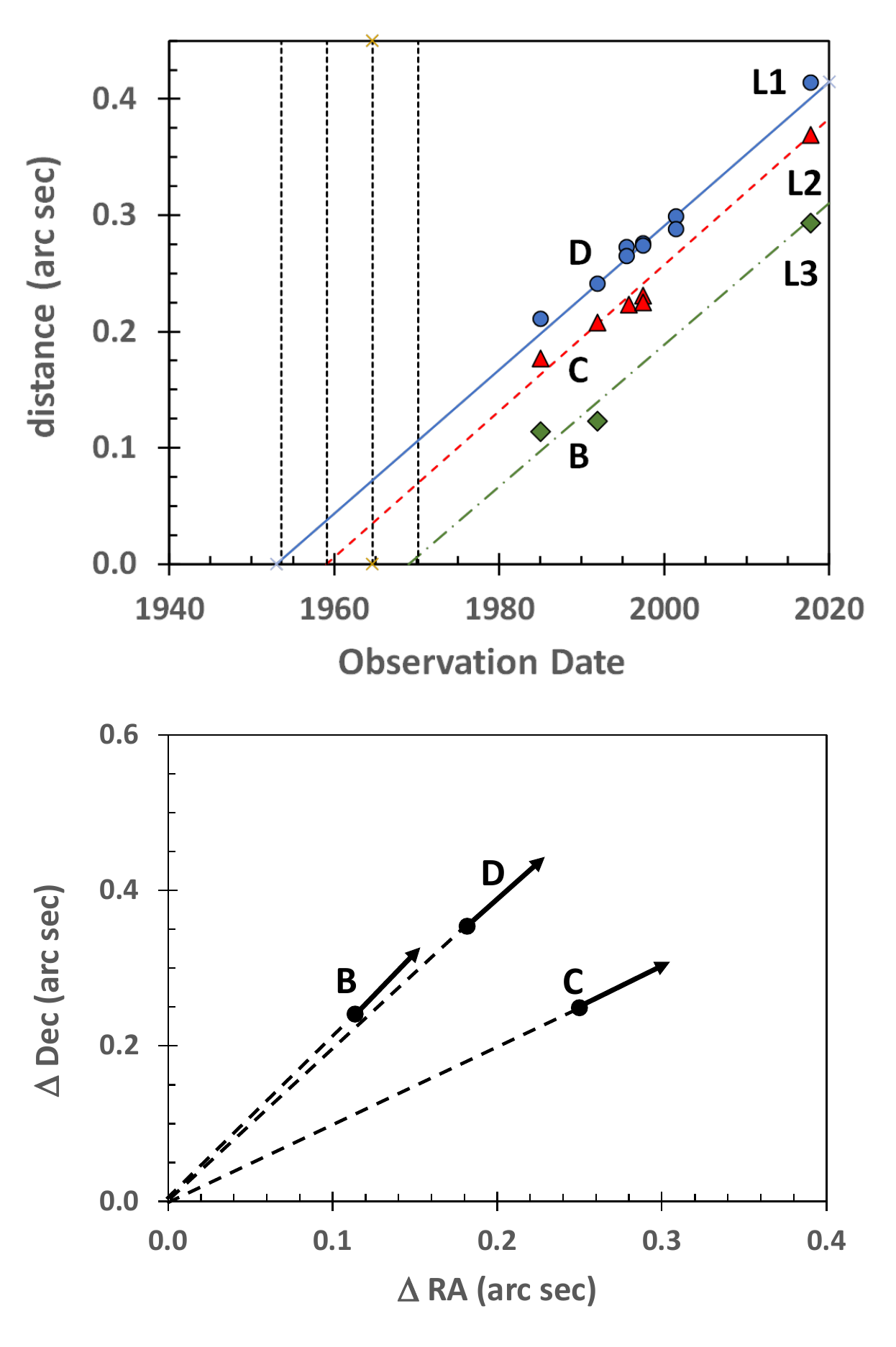}
\caption{Top: distance to the core  of Weigelt blobs B, C, and D  (L3, L2 and L1, respectively) as a function of time. Vertical lines correspond to epochs of minimum intensity in the high ionization lines. Bottom: position of the Weigelt blobs relative to the continuum compact source; arrows represent velocities}
\label{fig:Fig_16.pdf}
\end{center}
\end{figure} 
%%%%%%%%%%%%%%%%%%%%%%%%%%%%%%%%%%%%%%%%%%%%%%%%%%%%%%%%%%%%%%%%%%%%%%%%%%%%%%%
\begin{figure}
\begin{center}
\includegraphics[width=\columnwidth]{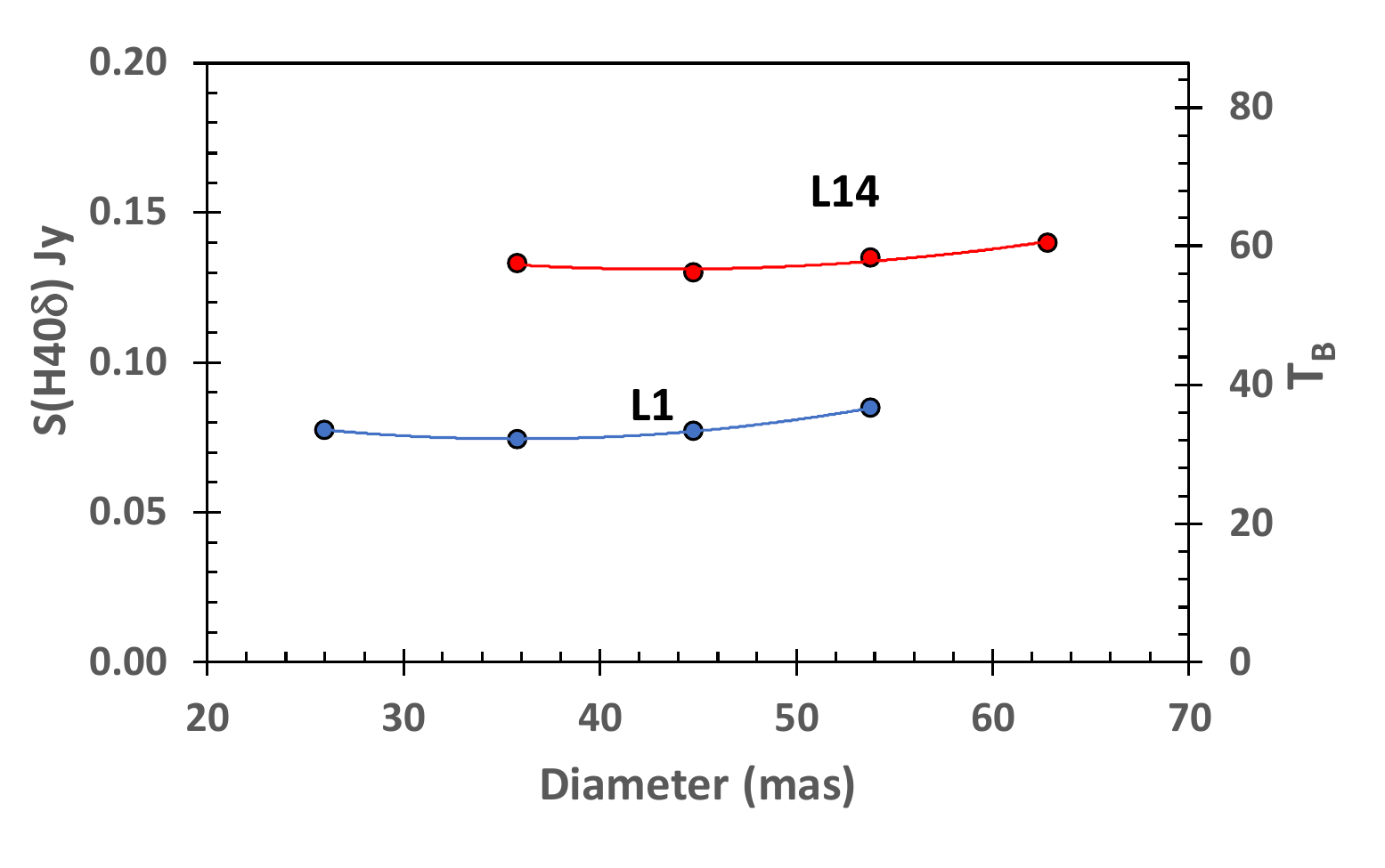}
\caption{Intensity of the H40$\delta$ line for the compact sources L1 and L14, for the same models than those in Figure \ref{fig:Fig_15.pdf} }
\label{fig:Fig_17.pdf}
\end{center}
\end{figure} 
%%%%%%%%%%%%%%%%%%%%%%%%%%%%%%%%%%%%%%%%%%%%%%%%%%%%%%%%%%%%%%%%%%%%%%%%%%%%%%%%%

Comparison of the integrated  continuum  emission with previous observations must take into account orbital phase and cycle. Considering the binary period of 2024 days  and the date 2014 July 1 (MJD 56850) as phase 13.0 \citep{dam19,cor17}, our ALMA Cycle~5 observations in 2017 Nov 20 corresponded to phase 13.61 and those of ALMA Cycle~0 \citepalias{abr14}, obtained in 2012 Nov 4, to phase 12.70. Taking into account the high eccentricity of the binary orbit ($\sim 0.9$), these two phases correspond to  similar orbital positions of the companion star, separated from apastron by $3^\circ$ and $5^\circ$, respectively. 
Therefore, both observations can be directly compared regarding the relative orbital position; the  total continuum flux density in the present observation is 32\% lower than that detected in 2012, when the source was not resolved by the ALMA $1\farcs52 \times 0\farcs75$  beam. 

While variability between continuum emission at different binary cycles can be expected, the lower flux density obtained in the present observations can be due to the existence of emission outside the maximum recovery scale  of the  ALMA configuration. 
This seemed to be also the case with  the ALMA 346~GHz continuum observations obtained by \citet{bor19} in 2016 October 24. The orbital phase at that time was 13.41 and the secondary orbital position  $13^\circ$ before apastron. The integrated flux density of the image was $30.5 \pm 1.2$ Jy, but the value raised to 38.9~Jy, compatible with the flux densities measured in 2012 and reported in \citetalias{abr14},  when a larger area was included in the integration.

 Figure \ref{fig:Fig_2.pdf} (center and bottom right images) shows the position of the 230 GHz continuum source relative to the $HST$ Wide Field Planetary Camera image, obtained with the filter F336W, centered at the wavelength of {3342 \AA}  \citep{mor98}. The position of several features described in the literature are indicated in the central image: the "fan" \citep{mor98}, the equatorial "skirt" \citep{smi98}, the "purple haze" \citep{smi04}, and the molecular torus \citepalias{smi18,bor19}.   The bottom image shows that the continuum radio source is located in the central part of the compact ultraviolet continuum source. The  upper left part of  Figure \ref{fig:Fig_2.pdf}, which will be discussed in Section \ref{subsec:other wavelenths}, presents the superposition of the 230 GHz contours on  the NB\_405 narrow filter image, which includes the Br$\alpha$ line, obtained with the VLT Adaptive Optics system NACO \citep{che05} . The position of the "Butterfly Nebula" is also indicate in this Figure.

%%%%%%%%%%%%%%%%%%%%%%%%%%%%%%%%%%%%%%%%%%%%%%%%%%%%%%%%%%%%%%%%%%%%%%%%%%%%%%%
\subsection{The H30\texorpdfstring{$\alpha$}{a} spectrum}
\label{subsec:spectra}

Figure \ref{fig:Fig_3.pdf} (top panel left) shows the H30$\alpha$  spectrum, centered at the line LSR frequency with velocity resolution of 1.262 km s$^{-1}$, integrated over the whole continuum emitting region. 
The central velocity of the narrow line is $-57.8$ km s$^{-1}$, its maximum flux density  $60.4 \pm 0.1$ Jy and HPW 45 km s$^{-1}$. The central velocity  is similar to that observed in 2012 \citepalias{abr14}, but the flux density is lower by 25\% and the line width larger by the same amount.
The weak  emission extends over several hundred km s$^{-1}$, both at positive and negative velocities.

A weak line with $3.9 \pm 0.1$ Jy maximum flux density   and $-176$ km s$^{-1}$ central velocity  is identified with the He30$\alpha$ line, 
separated from   H30$\alpha$  by $\Delta\nu / \nu = 0.000407$ or $-122$ km s$^{-1}$. Considering this frequency displacement, the actual velocity of the He atoms is $-54$ km s$^{-1}$, similar to the velocity of  the strong H30$\alpha$ line.

Figure \ref{fig:Fig_3.pdf} (top panel right) presents the spectrum of the compact region, centered on $\eta$ Car and integrated over an area of $0\farcs09$ radius. The narrow line, with maximum flux density of $2.64 \pm 0.01$ Jy is also centered at $-57.8$ km s$^{-1}$, blended with an even weaker line, centered at $\sim-100$ km s$^{-1}$; the He30$\alpha$ line is also present in the spectrum, as well as extended emission at larger negative and positive velocities. 
It is not clear if the narrow line belongs to the compact source or if it only represents  the  contribution of the superimposed extended NW region.

Figure \ref{fig:Fig_3.pdf} (bottom) displays the position velocity (PV) diagram of the region, collapsed along a line in the direction of the Homunculus axis (42\degr), with a width of 10~mas, which includes the position of the compact continuum source.
The position of the H30$\alpha$ and He30$\alpha$ lines are clearly visible in the NW direction, as well as the high positive and negative velocity features at both sides of the compact continuum  source position.

 The LSR velocity of $\eta$ Car, shown as a dotted horizontal line in  Fig. \ref{fig:Fig_3.pdf}, was estimated by \citet{smi04b} as $-19.6$ km s$^{-1}$  based on  the velocity of the H$_2$ lines in the walls of the  expanding   Homunculus. This velocity coincides with the  $-16.1 \pm 8.6$ km s$^{-1}$ mean velocity of the massive stars in the Tr16 cluster, which includes  $\eta$ Car \citep{kim18}. Similar velocities  (-15 to -20 km s$^{-1}$) are found in the CO$(1-0)$ lines of the Carina Molecular Complex, where  Tr16 is located \citep{reb16}.

The very good quality of our data enables us to make  images for each velocity channel. We separated therefore the H30$\alpha$ emission in narrow line, high negative and positive velocities, the $-100$ km s$^{-1}$ line, and extracted images for the He30$\alpha$ emission line. We  discuss the results  in the next subsections.
%%%%%%%%%%%%%%%%%%%%%%%%%%%%%%%%%%%%%%%%%%%%%%%%%%%%%%%%%%%%%%%%%%%%%%%%%%%%%%%
\subsection{The strong narrow H30\texorpdfstring{$\alpha$}{a} line}
\label{subsec:narrow}

 Figure \ref{fig:Fig_4.pdf} shows the iso-velocity intensity maps of channels with width of 1.262 km s$^{-1}$, showing the H30$\alpha$ narrow emission line, superimposed on the 230~GHz continuum emission image. The channels are separated in intervals of about 5 km s$^{-1}$, between -82 and -26 km s$^{-1}$.   The line emission at different velocities originates in different regions of the continuum image and  the compact  source, coincident with $\eta$ Car, does not contribute to the line emission at these velocities. 

Contour maps of the flux density,  integrated over the line profile between $-25$ and $-116$ km s$^{-1}$ (zero momentum), are presented in  Figure  \ref{fig:Fig_5.pdf}~(top), superimposed on  the the velocity weighted by flux density (first momentum) raster image. The regions of continuum and line enhancement coincide and  a velocity gradient in the north-west direction is clearly seen. 

The image of the H30$\alpha$ line at the velocity of maximum emission ($-54.6$~ km~s$^{-1}$) is comparable to that of the [\ion{Fe}{III}] line at similar velocity obtained with the {\it HST}/STIS,  presented by \citet{gul16}. The authors also reported variations of the  optical line velocity with position, although they had  less spatial resolution than in our ALMA data (i.e., $0\farcs1$ compared to 50~mas).  

More information is obtained from the  H30$\alpha$ line maximum flux density and the velocity of this maximum, which are presented as contours and raster images, respectively, at the bottom of  Figure \ref{fig:Fig_5.pdf}. This representation reveals that some of the enhanced emission regions in the zero momentum image are due to compact  not resolved components with very different velocities. 

We identified sixteen of these individual  components, labeled  L1 to L16; their spectra, integrated over a region of 66~mas radius are presented in Figure  \ref{fig:Fig_6.pdf}.
Most of them clearly show the superposition of more than one velocity component.

 The absolute position of these sources, continuum and peak flux densities of the H30$\alpha$ and He30$\alpha$ lines,  peak velocity of each line and HPW of the H line, projected distance $D$ to the compact continuum source and position angle $PA$ are presented in Table \ref{tab:Table_1}. The minimum HPW (17-18 km s$^{-1}$) of the lines that are not blended is of the order of the thermal velocity of a $10^4$ K  plasma. 

Also shown in Table \ref{tab:Table_1} are the emission measures $(EM)$, which will be discussed in Section \ref{subsec:parameters}. They were calculated from the continuum emission and from and H30$\alpha$ lines, assuming LTE conditions.

%%%%%%%%%%%%%%%%%%%%%%%%%%%%%%%%%%%%%%%%%%%%%%%%%%%%%%%%%%%%%%%%%%%%%%%%%%%%%%%

\begin{table*}

\caption{Properties of the strongest features embedded in the extended continuum and H30$\alpha$ emission region of the image shown in Fig. \ref{fig:Fig_5.pdf}}
\begin{center}
\label{tab:Table_1}
\begin{tabular}{ccccccccccccc}
\hline
{Comp.} &
{RA$^{a}$}&
{Dec$^{b}$} &
{$S$(cont)$^{c}$} & 
{$S({\rm H30}\alpha)$} &
{$V({\rm H30}\alpha)$} &{HPW} &
{$S({\rm He30}\alpha)$} &
{$V({\rm He30}\alpha)^{d}$} &
{$D^{e}$} &
{$PA$} &
{$\rm {EM(cont)}$} & 
{$\rm EM(H30 \alpha)$} \\
 &
 & 
{"} &
{Jy beam$^{-1}$} & 
{Jy beam$^{-1}$} & 
{km s$^{-1}$} &
{km s$^{-1}$} &
{Jy beam$^{-1}$} & 
{km s$^{-1}$} &
{"} &
{$\degr$} &
{$10^{10}$ \rm cm$^{-6}$ pc} &
{$10^{10}$ \rm cm$^{-6}$ pc} \\
\hline
C$^{f}$ & 3.532 & 3.980 & 1.13 &&  &  &  &  \\
L1$^{g}$ & 3.508 & 3.626 & 0.23 & 1.23 & -49.5 & 26 & 0.027 & -46`&  0.4 & 332.8 & 7.2  & 12.2 \\
L2$^{g}$ & 3.499 & 3.731 & 0.28 & 1.54 & -40.7 & 27 & 0.027 & -41 &  0.35 & 314.9 & 9.0  & 15.8 \\
L3$^{g}$ & 3.517 & 3.739 & 0.18 & 0.70 & -64.7 & 25 & 0.022 & -47 &  0.27 & 334.8 & 5.0  & 6.6  \\
L4 & 3.479 & 3.747 & 0.13 & 0.30 & -30.6 & 32 &       &     &  0.46 & 300.1 & 3.9  & 3.7  \\
L5 & 3.509 & 3.840 & 0.29 & 1.60 & -54.6 & 20 & 0.052 & -55 &  0.22 & 308.8 & 9.4  & 12.2 \\
L6 & 3.508 & 3.898 & 0.33 & 1.44 & -55.9 & 21 & 0.049 & -56 &  0.20 & 294.3 & 10.9 & 11.5 \\
L7 & 3.501 & 4.035 & 0.38 & 2.43 & -59.6 & 21 & 0.038 & -56 &  0.24 & 256.8 & 12.9 & 19.4 \\
L8 & 3.498 & 4.097 & 0.38 & 2.76 & -35.7 & 18 & 0.038 & -55 &  0.28 & 245.6 & 12.9 & 18.9\\
L9 & 3.491 & 4.046 & 0.22 & 1.43 & -58.4 & 20 & 0.052 & -60 &  0.32 & 258.0 &  6.9 & 10.9 \\
L10 & 3.475 & 4.011 & 0.22 & 0.60 & -29.3 & 33 & 0.030 & -30 & 0.43 & 265.9 &  6.9 &  7.5 \\
L11 & 3.528 & 3.782 & 0.43 & 1.86 & -54.6 & 30 & 0.056 & -45 & 0.20 & 351.3 & 15.1 & 21.2 \\
L12 & 3.539 & 3.813 & 0.40 & 2.70 & -34.4 & 21 & 0.045 & -56 & 0.18 &  17.6 & 13.8 & 21.6 \\
L13 & 3.540 & 3.879 & 0.34 & 1.92 & -60.9 & 21 & 0.061 & -60 & 0.12 &  31.0 & 11.3 & 15.3 \\
L14 & 3.526 & 3.918 & 0.38 & 2.51 & -52.1 & 17 & 0.058 & -60 & 0.08 & 323.8 & 12.9 & 16.2 \\
L15 & 3.532 & 3.988 & 0.59 & 2.25 & -55.8 & 21 & 0.070 & -56 & 0.01 & 359.0 & 23.1 & 18.0 \\
L16 & 3.516 & 3.941 & 0.29 & 1.06 & -62.1 & 22 & 0.056 & -56 & 0.13 & 287.8&  9.4 & 8.9 \\
\hline

\end{tabular}
\end{center}
{$a$}{ seconds from $10^{\rm h}45^{\rm m}$, uncertainty $0\fs001$}\\
{$b$}{ arc sec from $-59^\circ41'$, uncertainty $0\farcs001$\\}
{$c$}{ Estimated errors: S(cont) = 0.04 Jy/beam, S(H30$\alpha$)=0.04 Jy/beam, V(H30$\alpha$)=1 km~s$^{-1}$, HPW=2~km~s$^{-1}$, S(He30$\alpha$)=0.04 Jy/beam, V(He30$\alpha$)=1 km~s$^{-1}$\\}
{$d$}{ The HPW of the He30$\alpha$ line is not presented because it is blended with the negative velocity H30$\alpha$ line\\}
{$e$}{Distance to the compact continuum source\\}
{$f$}{ compact continuum source\\
{$g$}{The three first blobs, L1, L2, and L3 correspond to the Weigelt blobs D, C and B, respectively.} }
%\label{tab:Table_1}
\end{table*}
%%%%%%%%%%%%%%%%%%%%%%%%%%%%%%%%%%%%%%%%%%%%%%%%%%%%%%%%%%%%%%%%%%%%%%%%%%%%%%%

\subsection{The He30\texorpdfstring{$\alpha$}{a} line}
\label{sub:He30}

Figure \ref{fig:Fig_7.pdf} presents the  He30$\alpha$ line iso-velocity raster images. The $-56.3$ km s$^{-1}$  image, which correspond to a velocity of $-178.3$ km s$^{-1}$ at the frequency of the H30$\alpha$ line,  agrees very well with   the continuum image; the strongest emission arises from  compact components coincident with the compact continuum source, but it is not possible to assess if the emission is really part of the He30$\alpha$ transition or if it corresponds to the H30$\alpha$ transition with high negative velocity. 

The detection of He30$\alpha$ recombination lines coincident in position and velocity with both the continuum and narrow H30$\alpha$ line extended source  confirms  the presence of an  ionizing star of spectral type O or Of/WN7, with temperature of $\sim 37,000$~K, as suggested by \citetalias{ver05} and \citet{meh10}.

\ion{He}{ii} was detected  in the 4686 \AA~line that dramatically strengthens during periastron passage \citep{ste04,abr07,dav15,teo16}. 
At these epochs, the very fast equivalent width variability timescale ($\sim1$ week)  indicates that the optical emission comes from a compact source, not larger than $0\farcs5$, which coincides with the size of the He30$\alpha$ image.  
%%%%%%%%%%%%%%%%%%%%%%%%%%%%%%%%%%%%%%%%%%%%%%%%%%%%%%%%%%%%%%%%%%%%%%%%%%%%%%%
\subsection{High velocity H30\texorpdfstring{$\alpha$}{a} emission}
\label{subsec:high vel}

Figures \ref{fig:Fig_8.pdf} and \ref{fig:Fig_9.pdf} present the iso-velocity raster images of the H30$\alpha$ negative and positive high velocity emission, respectively, integrated over 14.6 km s$^{-1}$ velocity intervals, superimposed on the continuum contour map. The maximum flux densities between $-205$ and $-511$ km~s$^{-1}$, and 28 and 247 km~s$^{-1}$, are $0.43 \pm 0.09$~Jy beam$^{-1}$ at $-350$ km~s$^{-1}$ and $0.13 \pm 0.02$~Jy beam$^{-1}$ at 115 km~s$^{-1}$, respectively. 

The negative high velocity emission is concentrated in a compact region, to the south-east of the continuum source and has little structure, although it resembles the 40~mas Br$\gamma$ and continuum images of $\eta$ Car, obtained with the ESO Very Large Telescope Interferometer (VLTI) and the AMBER instrument \citep{wei16}.
 The positive high velocity emission arises from a weaker and more extended region, to  the north-west of the compact source. 
 
 Figure \ref{fig:Fig_10.pdf} shows contour maps of the positive and negative flux densities, integrated over the line profile, between velocities 22 and 250 km~s$^{-1}$ and -590 and -252 km~s$^{-1}$, respectively, superimposed on the continuum raster image. 
The positive and negative high-velocity emission seem to be aligned with the Homunculus axis.
This can be better seen in Figure \ref{fig:Fig_11.pdf}, where raster images of the H30$\alpha$ line peak flux density are shown for both negative and positive velocities, superimposed on the continuum contour map.    Inserts with the raster images of the line peak velocity, superimposed on contour maps of  the peak flux density show that for positive velocities there is a stratified  velocity distribution in the  north-west direction, with low  positive velocities in the region of the compact continuum source,  extending to 160 km s$^{-1}$ at the north end, including a narrow region in the middle, with $180-200$ km s$^{-1}$ velocities. The raster images of negative peak velocities do not present any conspicuous structure, being concentrated between $-300$ and $-350$ km s$^{-1}$. 

The combination of blue and red-shifted H30$\alpha$ emission defines the wind emission in the direction of the Homunculus nebula, that is, perpendicular to the plane of the binary orbit, probably coincident with the rotation axis of $\eta$ Car. 

   Similar velocity and spatial structure were observed by \citet{meh10} in  the [\ion{Fe}{iii}]    $\lambda  4659$  line, detected  with  $HST$/STIS in the period 1998 to 2004; the line  velocity  extends from $-250$  to $-400$ km~s$^{-1}$  at the position of $\eta$ Car, reaching $0\farcs1$ in the NE-SW direction and was attributed to the stellar wind at low latitudes. The lack of a corresponding redshifted emission was attributed to dust absorption. This structure is exactly what is  observed in the H30$\alpha$ recombination line except that the redshifted emission is present, although with lower intensity, probably  due to absorption by the optically thick free-free source surrounding $\eta$ Car, as discussed in Section \ref{Winds}
%%%%%%%%%%%%%%%%%%%%%%%%%%%%%%%%%%%%%%%%%%%%%%%%%%%%%%%%%%%%%%%%%%%%%%%%%%%%%%%
\subsection{The H bar}
\label{subsec:bar}

As mention in Section \ref{subsec:spectra}, the compact source line profile shows a weak  -100 km s$^{-1}$ component, blended with the stronger line centered at -58 km s$^{-1}$. The iso-velocity images presented in Figure \ref{fig:Fig_12.pdf} confirm the existence of a single source of emission, to the south-west of the compact continuum source, with velocities extending from -130 km s$^{-1}$ to lower negative velocities, reaching a maximum flux density of 0.6 Jy beam$^{-1}$ at -86 km s$^{-1}$, velocity at which it starts to  blend with other components of the extended north-west region. The -100 km s$^{-1}$ component can be followed up to lower velocities in Figure \ref{fig:Fig_4.pdf}, in a different flux density scale, until it disappears at -72 km s$^{-1}$.

This structure can be associated with the 1" length H$\alpha$ bar detected by \citet{fal96} using speckle imaging polarimetry. The sharp intensity drop found in the polarized optical line in the NW to SE direction was interpreted as absorption by a dusty equatorial disk around $\eta$ Car, with its rotation axis along the major Homunculus axis \citep{fal96}. However, since the H30$\alpha$ line is not affected by absorption, the asymmetry in the bar seems to be real.
%%%%%%%%%%%%%%%%%%%%%%%%%%%%%%%%%%%%%%%%%%%%%%%%%%%%%%%%%%%%%%%%%%%%%%%%%%%%%%%%%%%%%%%%%%%%%%%%%%%%%%%
\section{Discussion}
\label{Discussion}

The high angular resolution of the new ALMA data enabled to detect extended emission to the north-west of $\eta$ Car in the continuum and the H30$\alpha$ emission line, revealing a large number of compact, unresolved components with different velocities (Figs. \ref{fig:Fig_1.pdf},  \ref{fig:Fig_4.pdf} and \ref{fig:Fig_6.pdf}). 

 A strong unresolved continuum source is observed at the position of the binary system and high velocity H30$\alpha$ emission is detected in its spectrum in the direction of the Homunculus axis, probably associated with the aspherical wind of $\eta$ Car.
 
 Figure \ref{fig:Fig_13.pdf}  shows a schematic outlining the features detected in our images.
 In the next Sections we discuss the nature of the emission and the physical conditions of the different regions, as well as their implications on the different models discussed in the literature. 
 \subsection{The Central Source}
 \label{The Central Source}
 
 The detection of high energy emission by Fermi/LAT and H.E.S.S, and its variability along the  orbital period of $\eta$ Car \citep{rei15,ham18,hes20} rise the question of how much of the continuum radio emission is due to non-thermal synchrotron radiation.  
 
 To investigate this question, we calculate the thermal contribution of the ionized plasma, represented by the H30$\alpha$ line emission,  to the total continuum emission.
 
 First, the spectrum of the H30$\alpha$ line, integrated over a circle of  90 mas radius  centered on $\eta$ Car, was fitted by five Gaussian components that represent the negative and positive high velocity winds (G1 and G5), the He30$\alpha$ line (G2), the H bar (G3), and the narrow line, possibly resulting from the superposition of the NW source (G4).
 
 Table 2 presents the peak flux density, the central velocity and line width of each Gaussian component  and Figure \ref{fig:Fig_14.pdf} shows the observed spectrum, the fitted Gaussians and the residuals of the fitting.
%%%%%%%%%%%%%%%%%%%%%%%%%%%%%%%%%%%%%%%%%%%%%%%%%%%%%%%%%%%%%%%%%%%%%%%%%%%%%%%%%%%%%%%%%%%%%%%%%%%%%%%%% 
\begin{table}
\caption{Parameter of the five Gaussian components fitting the H30alpha emission line. }
\begin{center}
\label{tab:Table_2}
     \begin{tabular}{cccccc}
     \hline
     {Name} &
     {$V_{\rm L}$} &
     {$S_{\rm L}$} &
     {$\Delta v$} &
     {$EM$}
     \\
      & km s$^{-1}$ & Jy & km s$^{-1}$ & $10^{11}$ cm$^{-6}$pc \\
      \hline
   G1 & -332.5 & 1.34 & 217.6  & 1.02 \\
   G2 & -168.3 & 0.64 &  58.1  & 0.41 \\
   G3 & -93.8  & 2.03 &  57.8  & 0.53 \\
   G4 & -60.1  & 5.51 &  27.6  & 0.50 \\
   G5 &  83.9  & 0.56 & 254.0  & 0.13 \\
   \hline
     \end{tabular}
     \end{center}

 \end{table}
%%%%%%%%%%%%%%%%%%%%%%%%%%%%%%%%%%%%%%%%%%%%%%%%%%%%%%%%%%%%%%%%%%%%%%%%%%%%%%%%%%%%%%%%%%%%%%%%%%%%%%%%%%%% 
 
 The emission measure ${EM}_{\rm L}$ of each component, calculated from its line, is also presented in Table \ref{tab:Table_2}; it was obtained assuming a plasma with electron temperature $T_{\rm e}$ of $10^4$ K, optically thin to the H30$\alpha$ line radiation, which has a maximum brightness temperature $T_{\rm L}$:
 \begin{equation}
     T_{\rm L}=1.92 \times 10^3{\left(\frac{T_{\rm e}}{\rm K}\right)}^{-1.5}{\left(\frac{EM_{\rm L}}{\rm cm^{-6}pc}\right)}{\left(\frac{\Delta \nu}{\rm kHz}\right)},
 \end{equation}
where $\Delta \nu$ is the line width in kHz. The brightness  temperature $T_{\rm L}$ is related to the measured flux density $S_{\rm L}$ by:
 \begin{equation}
     S_{\rm L}=\frac{2kT_{\rm L}\nu_{\rm L}^2}{c^2}\Omega,
 \end{equation}
 \noindent
 where $\nu_{\rm L}$ is the central frequency of the line, $\Omega$ the solid angle of the source and $c$ the speed of light.The solid angle is calculated from:
 \begin{equation}
     \Omega=\pi \frac{\theta_{\rm maj}\theta_{\rm min}}{4 \ln 2},
 \end{equation}
 \noindent
 where $\theta_{\rm maj}$ and $\theta_{\rm min}$ are twice the major and minor axis of the ellipse that represent the source (in our case both  equal to 180 mas).
 
 The emission measure necessary to explain the continuum emission, $EM_{\rm c}$ assumed to be of free-free origin, is obtained from the opacity $\tau$:
\begin{equation}
 \tau=3.014 \times 10^{-2}\left(\frac{T_{\rm e}}{\rm K}\right)^{-1.5}\left(\frac{\nu}{\rm GHz}\right)^{-2}\left(\frac{EM_{\rm c}}{\rm cm^{-6}pc}\right)<g_{ff}> 
 \end{equation}
 \noindent
 with
  \begin{equation}
 <g_{ff}>=\ln \left[4.955\times 10^{-2} \left(\frac{\nu}{\rm GHz}\right)^{-1}\right]+1.5\times \ln \left(\frac{T_{\rm e}}{\rm K}\right)
 \end{equation}
The opacity $\tau$ is obtained from the observed flux density $S_{\rm c}$:
 \begin{equation}
     S_{\rm c}=\frac{2kT_{\rm e}\nu^2}{c^2}\left(1-e^{-\tau}\right) \Omega.
 \end{equation}
 
 From Table \ref{tab:Table_2} we find that the total emission measure necessary to account  for the line intensity is $2.59\times 10^{11}$ cm$^{-6}$pc, while that necessary to explain the continuum is $2.46\times 10^{11}$ cm$^{-6}$pc, showing that all the continuum emission can be explained by thermal processes, at least during the orbital phases in which the stars are farther apart. During the 2003.5 periastron passage, \citet{abr05} found, in single dish daily observations,  a 1 Jy increase in the 7~mm continuum lasting for a week, which they attributed to  emission from the shocked region formed by the colliding winds. More high resolution data at different orbital phases of the binary system are needed to confirm this assumption.
 %%%%%%%%%%%%%%%%%%%%%%%%%%%%%%%%%%%%%%%%%%%%%%%%%%%%%%%%%%%%%%%%%%%%%%%%%%%%%%%%%%%%%%%%%%%%%%%%%%%%%%
 \subsection{The high velocity winds}
\label{Winds}

 The contribution of the $\eta$ Car wind to the total continuum emission can be obtained by dividing the emission measure calculated from its spectral lines G1 and G5, presented in Table \ref{tab:Table_2}, by the  emission measure derived from the total spectrum. 

Even if we exclude the strong narrow line, the remaining H30$\alpha$ line profile of the compact source  differs from the $0\farcs15$ resolution H$\delta$ profiles obtained by \citet{nie07} with the {\it Hubble Space Telescope} Space Telescope Imaging Spectrograph ({\it HST}/STIS) during a complete binary cycle,  which show  smooth lines centered at approximately the systemic velocity of $\eta$ Car and extending to $\pm 500$ km s$^{-1}$. 
As the continuum spectrum reported in \citetalias{abr14} presents characteristics of a compact \ion{H}{II} region that becomes optically thin ($\tau \sim 1$) around 300~GHz, the asymmetry in the high velocity H30$\alpha$ spectrum can be attributed to absorption, if part of the material that produces the positive velocity  line is behind the continuum source.
 Therefore, we consider that the total wind emission is twice the value of the  blueshifted component flux density and compare it to the theoretical value  expected from free-free emission of an ionized wind \citep{pan75}:

\begin{equation}
\begin{split}
    S_{\rm \nu}({\rm wind}) & =5.12\left[\frac{\nu}{10~ {\rm GHz}}\right]^{0.6}\left[\frac{T_{\rm e}}{10^4 {\rm K}}\right]^{0.1}\left[\frac{\dot{M}}{10^{-5} {\rm M_{\odot} yr^{-1}}}\right]^{4/3} \\ &  ~~~~~~~~~~~~~~~\times\left[\frac{v_{\infty}}{10^3~{\rm km s^{-1}}}\right]^{-4/3}\left[\frac{d}{\rm kpc}\right]^{-2}
    \end{split}
\end{equation}
Using $10^{-3}$ M$_{\odot}$ yr$^{-1}$ for the mass loss rate $\dot{M}$ and 500 km s$^{-1}$ for the terminal wind velocity $v_{\infty}$ we obtain $S_\nu({\rm wind})=7.4$~Jy.
The value derived from the observations is 3.7 Jy, smaller than the theoretical value. The reason is  that because of the fast rotation of $\eta$ Car, its wind  is aspherical and the mass loss rate and velocity depend on latitude $\theta$ \citep{dwa02} as:

\begin{equation}
    \dot{M}(\theta)=\dot{M}(90\degr) \left[1-\omega^2\cos^2(\theta)\right]
\end{equation}
\begin{equation}
    v_{\infty}(\theta)= v_{\infty}(90\degr) \left[1-\omega^2\cos^2(\theta)\right]^{1/2},
\end{equation}
\noindent
$\omega$ is the rotation velocity of $\eta$ Car relative to the critical value of $({\rm G}M/R^3)^{1/2}$, where $G$ is the gravitational constant and $M$ and $R$ its mass and radius.

Taking into account that the wind is not isotropic, the observed terminal  velocity is  the projection of the real terminal velocity into the line of sight. Assuming that the rotation axis of $\eta$ Car coincides with the Homunculus axis, the true terminal velocity  is  $\sim 700$~km s$^{-1}$, compatible with the values measured by \citet{smi03}. The effect of including the latitude dependence of the mass loss rate and wind velocity assuming $\omega=0.9$ is to decrease the expected flux density by 67\%, while the increase in the terminal velocity will decrease it by 63\%, resulting in an expected flux density from the wind of 3.2 Jy, in agreement with the 3.7 Jy derived from the observations.

The radius $R_{\rm v}$ at which half of the wind continuum radiation is emitted was also derived by \citet{pan75}:
\begin{equation}
\begin{split}
    R_{\rm \nu}({\rm wind}) & =6.23\times 10^{14}\left[\frac{\nu}{10~ {\rm GHz}}\right]^{-0.7}\left[\frac{T_{\rm e}}{10^4 {\rm K}}\right]^{-0.45} \\ &
    ~~~~~~~~~~~~\times\left[]\frac{\dot{M}}{10^{-5} {\rm M_{\odot} yr^{-1}}}\right]^{2/3}   \left[\frac{v_{\infty}}{10^3~{\rm km s^{-1}}}\right]^{-2/3} {\rm cm}
    \end{split} 
\end{equation}

Using this expression we find the extension of the wind in the directions of the Homunculus axis is  $\sim 130$ AU or $\sim 60$ mas, in agreement with the observations presented in Figure \ref{fig:Fig_10.pdf}.

The critical value of the mass loss rate that can be fully ionized by  $\eta$ Car was calculated by \citet{mor98} as $3.3\times 10^{-4} {\rm M}_\odot$ yr$^{-1}$; since the value inferred from the spectra is smaller than that attributed to $\eta$ Car, the  ionizing front must be trapped inside the wind. Although it is possible that the dense base of the wind is ionized by collisions  or by Balmer continuum photons, as discussed  by \citet{nat91}, it is still necessary to add the ionizing photons of the companion star to account for the ionization of the extended NW region. 
This raises the question of the position of the companion star relative to $\eta$ Car during the ALMA observations. Although the observations were obtained close to apastron, single dish observations during the complete orbital period show free-free emission during all the cycle, except for a few month during periastron passage, implying that the ionizing photons reach the region during all this time. The same behavior was observed at lower frequencies and larger distances by \citet{dun03}, meaning that the ionizing front reaches also the little Homunculus and maybe the surrounding torus or disk. For these reasons we believe that the relative position of the stars along the orbit cannot be  fully derived from the ALMA data, but can be better determined by the observed and modeled position of the fossil material left by the wind-wind collision zones.
 %%%%%%%%%%%%%%%%%%%%%%%%%%%%%%%%%%%%%%%%%%%%%%%%%%%%%%%%%%%%%%%%%%%%%%%%%%%%%%%%%%%%%%%%%%%%%%%%%%%%%%%%
 \subsection{The physical parameters of the {\bf}NW compact regions }
\label{subsec:parameters}

 The spectrum of the NW extended region is dominated by the very strong and narrow H30$\alpha$ recombination line. Previous non-resolved ALMA observations \citepalias{abr14} of the recombination lines H42$\alpha$, H40$\alpha$, H30$\alpha$,  H28$\alpha$, and H21$\alpha$, and their respective continuum were modeled as a homogeneous expanding ionized shell of about $0\farcs2$ radius and $0\farcs02$ width, with an electron density $n_e = 1.3\times 10^7$ cm$^{-3}$ and a temperature $T_e = 1.7\times 10^4$ K. 

The size of the shell agrees with the present observations and its width is smaller than the HPBW, but the velocity distribution shown in Figure \ref{fig:Fig_6.pdf} is not compatible with that of an expanding shell; instead the velocities of the different compact components seem to follow a random pattern. 

 In Table \ref{tab:Table_1} we present the $EM$s of the compact components inferred  from the line profiles assuming LTE conditions and  their sizes equal to the beam size, and also the $EM$s inferred from the underlying free-free continuum; equations 1-6 were used in the calculations. In most of the sources, the $EM$ derived from the lines is larger than that derived from the continuum, a clear sign of NLTE.

To investigate the physical conditions of the observed individual non resolved sources, we assume that they are homogeneous ionized spheres, with diameter $D$ smaller than the HPBW. We solved the NLTE transfer equations as in \citetalias{abr14} and determined the combination  of $T_e$, $n_e$ and $D$ values necessary to match simultaneously the observed continuum flux density and maximum line intensity, as presented in Table \ref{tab:Table_1}. We modeled two components, L1 and L14, because each one of them seems to arise from a single velocity region. We first selected a value for the diameter and three values for $T_e:$ (1.0, 1.1 and 1.2)$\times 10^4$ K, and determined the values of $n_e$ necessary to match the observed continuum flux density, obtaining the values of the line maximum flux density. We then obtained, by interpolation, the values of $T_e$ and $n_e$ for which the model and observed line flux density coincide. We repeated the procedure for different values of the source diameter; the results are presented in Figure \ref{fig:Fig_15.pdf}, with the combinations of $n_e$ vs. $D$ in the top panel, $T_e$ vs. $n_e$ in the middle and the line amplification, represented by the ratio of the NLTE to LTE maximum line intensity in the bottom. The values of $n_e$ are compatible with what was found in \citetalias{abr14} although $T_e$ is somehow smaller, more compatible with the temperature expected in a compact \ion{H}{II} region.

For both sources and for all the physical conditions presented in Figure \ref{fig:Fig_15.pdf}, the turnover frequency of the continuum spectrum (where $\tau \sim 1$) is very close to the frequency of observation. This is consistent with the spectrum between 90 and 670 GHz measured with ALMA during Cycle~0 \citepalias{abr14}.

The mass of the model sources vary between 0.9 and 3.0$\times 10^{-5}$ M$_\odot$ for L1 and 1.8 and 4.0$\times 10^{-5}$  M$_\odot$  for L14, for increasing values of $D$. Considering that all the compact sources have similar masses, we obtain a total mass for the extended region of $10^{-4} - 10^{-3}$ M$_\odot$. A similar result was obtained using the mean density of the extended region necessary to reproduce the observed continuum flux density ($n_e = 2.8 \times 10^6$ cm$^{-3}$ for $D=1.8\times 10^{16}$ cm).

%%%%%%%%%%%%%%%%%%%%%%%%%%%%%%%%%%%%%%%%%%%%%%%%%%%%%%%%%%%%%%%%%%%%%%%%%%%%
\subsection{Comparison with observations at other wavelengths}
\label{subsec:other wavelenths}
The detection of a large number of compact (< 60 mas) sources within $0\farcs6$ to the north-west of $\eta$ Car, emitting in the 230 GHz continuum and in the H30$\alpha$ and He30$\alpha$ recombination lines,  with densities and temperatures of the order of $10^7$ cm$^{-3}$ and $10^4$ K, as estimated in the previous subsection,  must be analyzed in the context of previous observations of the same region at different wavelengths and with similar resolutions.

The region, labeled by \citet{che05} the "Weigelt Complex", was intensively studied since the discovery, using speckle interferometry techniques, of compact sources (< 30 mas) at distances smaller than $0\farcs3$ from $\eta$ Car \citep{wei95,hof88}. The brightest sources besides source A, identified with the star, were labeled "Weigelt blobs" B, C, and D; they present narrow permitted lines of H, He and \ion{Fe}{ii}, and forbidden lines of [\ion{N}{i}], [\ion{Fe}{ii}], [\ion{Ne}{iii}], [\ion{Fe}{iii}], and other ions, with heliocentric velocities in the range of -45 to -48 km s$^{-1}$ and  FWHM of about 22 km s$^{-1}$ \citep{dav95,dav97}. 
The UV to IR spectra of the Weigelt blobs is also affected by resonance fluorescence, occurring specially in the \ion{Fe}{ii}  lines, but also in \ion{O}{i}, \ion{Cr}{ii}, \ion{Fe}{iii}, and \ion{Ni}{ii}. In these ions, the upper energy levels are overpopulated because of energy coincidence with \ion{H}{i} Ly$\alpha$ photons \citep{joh00,joh01,joh05}. The existence of these transitions  imply that the Ly$\alpha$ photons must be local.

The physical conditions of the regions that present spectra of low ionization elements was estimated from emission models as $n_e\sim10^6$ cm$^{-3}$ and $T_e \sim 7000$~K \citep{ver02}.
Models of spectra of high ionization elements and high excitation lines, on the other hand, require photoionized regions of H$^+$ and He$^+$ with  densities  and temperatures of $10^7$ cm$^{-3}$ and $10^4$ K \citepalias{ver05,meh10}. 
The narrow line spectra of the high ionization elements show also strong dependence on the phase of the binary orbit, disappearing completely during the spectroscopic events, while the lines of the low ionization elements, like those of [\ion{Fe}{ii}] remain visible, and are even enhanced \citep{joh00,smi04,har05,nie07,meh10,ham12,rem13,gul16}.

The nature of the Weigelt blobs was explained by different models, which were modified as new observational evidences became available. The first models assumed that the blobs are formed by neutral H, extending from the ionizing O type supergiant or  WR star to distances of $0\farcs3$ or more; as a consequence, the high ionization elements are found in an extended region close to the ionizing source while the low ionization material populates a narrow region at  the surface of the blobs \citep{ver02,joh01}. This model was contested by \citetalias{rem13}, because the ions with higher ionization energy were observed  at larger distances from the star. A new qualitative model, similar to that proposed by \citet{smi04}, assumed that the blobs are dense condensations of low ionization material, with its surface  ablated by photoionization, and that the evaporated material, now highly ionized, moves outward around the condensation, accelerated by radiative forces. 

Another constrain for the models came from the infrared observations obtained with the ESO/VLT Adaptative Optics system NACO and narrow band filters centered at the wavelengths of the H Pf$\gamma$ and Br$\alpha$ lines, which  showed Weigelt  blobs C and D clearly resolved. These observations were interpreted as  evidence of the existence of dust inside the blobs \citep{che05}.  However, comparison of the 3.74 $\mu$m (Pf$\gamma$) image with red continuum images of the blobs presented by \citet{mor98} did not show perfect  spatial coincidence at both wavelengths, which raised the  possibility that the blobs are part of a larger region that become visible only where the dust absorption is lower.  In fact, dust grains can condensate at the location of the Weigelt blobs \citep{fal96}. \citet{dav97} estimated the grain temperature at this site to be $\sim 1000$ K. However, the presence of the  $10^4$ K source detected by ALMA will probably destroy the grains, if present.

The ALMA observations reported in this paper show the existence of compact fully ionized regions, with densities and temperatures similar to those necessary to explain the spectra of low and high ionization elements in the Weigelt Complex, as well as the high excitation lines. Their radial velocities are also similar to  those of the  UV-IR lines, considering  that the former are relative to the LSR while the latter are heliocentric, and there is a difference of -11.6 km s$^{-1}$ between them. All these considerations suggest  the identification of the compact radio sources with the UV to IR blobs in the Weigelt Complex.

It should be pointed out that it is the high velocity resolution of the ALMA observations that allowed the identification of compact sources, otherwise, they would appear as larger condensations of matter with a wider velocity profile, as observed by \citetalias{rem13}. In a similar way, the UV to IR spectra of the  blobs show that the lines of different elements originate in different positions and can have different  velocities \citepalias{smi04}, and therefore are  probably formed in  different compact radio sources. 

There is still a question that remains to be treated: the existence of dust in the center of the Weigelt blobs, as suggested by the IR observations of \citet{che05}.
Analysis of these observations show that the Weigelt blobs were only conspicuous in the narrow filters with central wavelengths coincident with the hydrogen lines Pf$\gamma$ and Br$\alpha$; the  Pf$\gamma$ line was calibrated, resulting in a flux density of $520 \pm 70$ Jy for the central source, identified as $\eta$ Car, and  lower flux densities for the other detected blobs. Based on the ALMA observations, we claim that the observed emission comes from the Pf$\gamma$ H line instead of dust. To support this assumption we calculated the expected Pf$\gamma$ line flux density $S(\rm{Pf}\gamma$) of one of our typical compact sources, by scaling the emissivity coefficients $\epsilon$ of the H30$\alpha$ and Pf$\gamma$ transitions, for ionized H density of $10^7$ cm$^{-3}$ and temperature $10^4$ K given by \citet{sto95}. The resulting flux density of the Pf$\gamma$ line is given by:
\begin{equation}
S(\rm{Pf}\gamma)=\frac{\epsilon(\rm{Pf}\gamma)}{\epsilon(\rm{H30}\alpha)}\frac{\nu(\rm{H30}\alpha)}{\nu(\rm{Pf}\gamma)} \it{S}(\rm{H30}\alpha)
\end{equation}
The ratio of the emissivity coefficients is $8.7\times 10^3$, and the flux density of the H30$\alpha$ line is $S(\rm{H30}\alpha)\sim 2$ Jy, for a source of 50 mas diameter, resulting in an expected flux density of  47 Jy for the Pf$\gamma$ emission of the compact radio source. Considering that there can be a superposition of sources in the 100 mas beam of the IR observation, we find $S(\rm{Pf}\gamma)=190$ Jy for the IR blob, showing that the observed IR emission can be explained by H line emission, and that the presence of dust emission is not required. 

 A complementary argument is obtained from the excellent agreement between the structures seen in the 230 GHz continuum and in the Br$\alpha$ images in Figure \ref{fig:Fig_2.pdf}. Since the images obtained with wider filters, which are dominated by dust emission, do not show these structures, we conclude that the dust is located between the compact regions and the observer, shielded from the UV radiation necessary to sustain the ionized plasma detected in the ALMA observations. The existence of this gas layer is corroborated by the absorption of CO and HCN in front of the continuum source detected  by  \citet{bor19} at 345 GHz.
%%%%%%%%%%%%%%%%%%%%%%%%%%%%%%%%%%%%%%%%%%%%%%%%%%%%%%%%%%%%%%%%%%%%%%%%%%%
\subsection{The Weigelt blobs}
\label{Weigelt}
 In the previous subsection we identified the compact radio sources with the UV-IR blobs seen in the Weigelt Complex.
 
 The origin and evolution of these blobs  is not clear; \citet{dav97}, \citet{smi04} and \citetalias{dor04}, studied the proper motions of Weigelt blobs C and D, and attributed them to ejections that occurred some time between 1910 and 1941. 

In order to determine more accurately the ejection epochs of the blobs and their proper motions, we used their measured positions at previous epochs  to estimate their probable position at the epoch of our observations. We were able to identified blobs C and D with the compact sources L2 and L1, with position angles $315\degr$ and $333\degr$, respectively. We tentatively identified Weigelt blob B with L3, although it was only present in the \citet{wei86} observations. Their positions relative to $\eta$ Car are shown in Figure \ref{fig:Fig_16.pdf}.
The velocities, obtained from a linear fitting of their positions are $6.2 \pm 1.6$ mas yr$^{-1}$, $6.0 \pm 1.6$ mas yr$^{-1}$ and $5.7 \pm 2.3$ mas yr$^{-1}$, for blobs D (L1), C (L2) and B (L3), respectively. The  linear velocity corresponding to an angular velocity of 6 mas yr$^{-1}$
is 67 km~s$^{-1}$, similar to the Weigelt blobs radial velocities. The ejection epochs obtained from the linear fitting are 1952.6, 1957.1 and 1967.6, respectively. In the upper part of Figure \ref{fig:Fig_16.pdf} we also show the epochs of minimum in the high excitation line intensities, presumably related to  periatron passage: 1953.6,  1959.1,  1964.6 and  1970.2, corresponding to cycles 2, 3, 4 and 5 according to \citet{dam08}, considering that  the first spectroscopic event was reported by
\citet{gav53}.   In the lower part of the Fig. \ref{fig:Fig_16.pdf} we show the position of the Weigelt blobs relative to the central continuum source, with vectors indicating the direction and the relative magnitude of their velocities in the plane of the sky. 
We conclude that the Weigelt blobs were ejected close to periastron passage during the firsts spectroscopic events. It is possible that the other compact sources observed in the H30$\alpha$ line image, closer to $\eta$ Car, were ejected later, and if so, their evolution should be measured in future observations.  

From the radial velocities and proper motions, we calculate the angles of the D, C and B blob's trajectories relative to the plane of the sky as ${35\fdg0}{^{+10\degr}_{-6\degr}}$, ${31\fdg2}{^{+9\degr}_{-6\degr}}$ and ${45\fdg5}{^{+14\degr}_{-9\degr}}$, all of them compatible with the position of the equatorial plane of the Homunculus nebula.

 The observed angular velocities are larger than those derived previously by other authors \citepalias{dav97,smi04,dor04}; the reason is the larger time span  in our observations. Of course, this result depends on the accuracy of the identification of the Weigelt blobs with the compact sources detected by ALMA, but unless the blobs had already vanished, the only  sources in the data with position angles and distances compatible with those of the blobs are L1, L2 and L3. 
%%%%%%%%%%%%%%%%%%%%%%%%%%%%%%%%%%%%%%%%%%%%%%%%%%%%%%%%%%%%%%%%%%%%%%%%%%%%%%%%%%%%%%%%%%%%%%%%%%%%%%%%%%
\subsection{HCO+ or H40\texorpdfstring{$\delta$}{d} emission line?}
\label{sec:HCO+}
%%%%%%%%%%%%%%%%%%%%%%%%%%%%%%%%%%%%%%%%%%%%%%%%%%%%%%%%%%%%%%%%%%%%%%%%%%%%%%%%%%%%%%%%%%%%%%%%%%%%%%%%%%%
\citet{bor19} reported ALMA observations of CO, HCN and HCO$^+$ in the direction of $\eta$ Car.
The molecular gas, as traced by the CO(3-2) and HCN(4-3) emission lines, is seen in a disrupted 2" torus surrounding the binary system (see also Smith et al. 2018). In contrast, the HCO$^+$(4-3) emission line is also detected in an asymmetric, extended structure north-west of the star, that coincides precisely with that of the 230 GHz continuum and H30$\alpha$ line emission described in this paper. This spatial coincidence and the fact that the presence of HCO$^+$ is unexpected in a highly ionized gas led us to explore further the nature of this emission line and see if it could be instead a hydrogen recombination line. 

The rest frequency of the HCO$^+$(4-3) line is 356.734 GHz; the nearest hydrogen recombination line is H40$\delta$, with a rest frequency of 356.658~GHz. Considering the difference in frequencies, if the line observed by \citet{bor19} were H40$\delta$, the corresponding velocity would be -63.7~km~s$^{-1}$, coincident with the velocity of the extended region. This coincidence strongly suggests that the 356.7 GHz emission line, outside the 2" torus is the H40$\delta$ recombination line. 

To support this suggestion, we calculated the intensity of the H40$\delta$  line in the compact sources L1 and L14, as it was done for the H30$\alpha$ line. The resulting flux densities are presented in Figure \ref{fig:Fig_17.pdf} as a function of the source diameter, for the same combination of electron densities and temperatures  necessary to fit the 230 GHz continuum and H30$\alpha$ lines.
The brightness temperatures are also presented in the figure, showing that  the profile of the  line detected by \citet{bor19} can be explained by  the superposition of the line profiles of the compact sources with different velocities as detected in the H30$\alpha$ line.

Although the calculations were made assuming NLTE, the line intensities show that the population is in LTE.
We can notice the similarity of the -54.6 km~s$^{-1}$ image in our Figure \ref{fig:Fig_4.pdf} with that of Figure 4 of \citet{bor19}. The only difference is the intensity of components  L14 and L15, which are much stronger in the H30$\alpha$ line. This is consistent with the fact that those lines are amplified by NLTE effects in the H30$\alpha$ lines.

Based on the spatial coincidence of the emissions of the H30$\alpha$ and the 356.7 GHz emission line reported by \citet{bor19} and the results of the above analysis, we conclude that the emission line centered at the position of the binary system is the H40$\delta$ recombination line. 
%%%%%%%%%%%%%%%%%%%%%%%%%%%%%%%%%%%%%%%%%%%%%%%%%%%%%%%%%%%%%%%%%%%%%%%%%%%%%%%%%%%%%%%%%%%%%%%%%%%
\section{Conclusions}
\label{sec:Conclusions}

In this paper, we report on new ALMA high-angular resolution observations of the H30$\alpha$ and He30$\alpha$ recombination lines and the underlying continuum of the $0\farcs8 \times 0\farcs8$ central region of $\eta$ Car. The unprecedented angular resolution of  50 mas (or 110 AU) allowed us to explore the extended emission north-west of the binary star tracing the most recent mass-loss history. The main findings based on these new observations are as follows.  

1. The 230 GHz continuum, H30$\alpha$ and He30$\alpha$ recombination line images of $\eta$~Car were resolved with the $0\farcs065\times0\farcs043$ beam obtained using ALMA in the extended configuration.
The continuum image consists of a structureless core  of $0\farcs12$ HPW and weaker emission extending over $\sim0\farcs6$ in the NW direction that presents several regions of enhanced emission. 

2. The spatially integrated spectrum is similar to that obtained in \citetalias{abr14}, with a strong H30$\alpha$ narrow line centered at  $-58$~km~s$^{-1}$ and much weaker extended emission at higher negative and positive velocities. The narrow line is much weaker in the spectrum integrated over the compact continuum source. The He30$\alpha$ line is also detected in both spectra.

3. The high  quality ALMA data allowed us to extract images of the H30$\alpha$ line emission with 1.262 km s$^{-1}$ velocity resolution, revealing that  maximum emission in different regions occurs at different velocities. Sixteen compact sources with well defined velocities were identified; their spectra, when not blended, have HPW of $\sim20$~km~s$^{-1}$, compatible with the thermal velocity of  $10^4$~K plasma. Three of these sources (L3, L2, and L1) were identified with Weigelt blobs B, C, and D, respectively.

4. The H30$\alpha$ recombination line profiles of the non-resolved sources and their underlying continuum flux densities  were reproduced assuming homogeneous spheres of ionized gas in NLTE, with densities and temperatures  characteristic of compact \ion{H}{ii} regions ($10^7$~ cm$^{-3}$ and $10^4$~K, respectively). These compact sources probably represent the UV to IR blobs observed in the Weigelt Complex. 

5. The high negative velocity line emission arises from a compact source  SE  of $\eta$~Car, while the positive velocity emission extends in the NW direction. Both seem to be aligned with the Homunculus axis and to arise from the stellar wind enhanced in the polar axis of the fast rotating $\eta$ Car.

6. We also detected concentrated emission at the velocity of about $-100$ km~s$^{-1}$, corresponding to the "bar" in the H$\alpha$ speckle polarimetric observations of \citet{fal96}.

7. For the three  compact sources in the H30$\alpha$ images that  correspond to the  Weigelt blobs D, C and B we determined the proper motions  to be about 6 mas~yr$^{-1}$ and their ejection epochs 1952.6, 1957.1 and 1967.6, respectively, close to the epochs of periastron passage. 

8. The striking similarity between our images of the extended region, in the continuum and in the narrow H30$\alpha$ line, and those of the  345 GHz continuum and the emission line reported by \citet{bor19} that they attributed to HCO$^+$(4-3), led us to investigate the possibility that the observed line is instead a  recombination line. 
The frequency of this line is indeed very close to that of the H40$\delta$ recombination line and, would in this case, be at a velocity of -64 km~s$^{-1}$, which corresponds to the velocity of the NW extended emission region. Based on this identification, we modeled the compact regions detected in our observations and were able to reproduce the intensity of the line observed by  \citet{bor19}, supporting our interpretation that this emission line is indeed H40$\delta$.

Future high-angular resolution observations using ALMA will be able to measure the proper motions of all the compact sources detected in this work, and also the effect of periastron passage on their continuum flux density and spectra.

\section{Data Availability}
The data underlying this article are available in ADS/JAO.ALMA$\#$2017.1.00725.S

\section*{Acknowledgements}

We would like to thank our referee, John Bally, for insightful comments that helped to improve this paper.
This paper makes use of the following ALMA data: ADS/JAO.ALMA$\#$2017.1.00725.S. ALMA is a partnership of ESO (representing its member states), NSF (USA) and NINS (Japan),
together with NRC (Canada), MOST and ASIAA (Taiwan), and KASI (Republic of Korea), in
cooperation with the Republic of Chile. The Joint ALMA Observatory is operated by
ESO, AUI/NRAO and NAOJ.
ZA and PPB acknowledge Brazilian agencies FAPESP (grant \#2011/51676-9 and \#2014/07460-0) and CNPq (grant \#305768/2015-8)
%%%%%%%%%%%%%%%%%%%%%%%%%%%%%%%%%%%%%%%%%%%%%%%%%%%%%%%%%%%%%%%%%%%%%%%%%%%%%%%%%%%%%%%%%%%%%%%%%%%%%%%%%%

%%%%%%%%%%%%%%%%%%%%%%%%%%%%%%%%%%%%%%%%%%%%%%%%%%

% Don't change these lines
\bsp	% typesetting comment
\label{lastpage}
\end{document}